\documentclass[english,preprint,aps,prd,showpacs,superscriptaddress,nofootinbib,tightenlines]{revtex4}
\usepackage[T1]{fontenc}
\usepackage[latin9]{inputenc}
\usepackage{float}
\usepackage{amsmath}
\usepackage{graphicx}
\usepackage{esint}
\usepackage{ulem}
\usepackage{amsfonts}
\usepackage{multirow}
\usepackage{mathrsfs}
\usepackage{graphicx}
\usepackage{amsmath}
\usepackage{amssymb}
\usepackage{bm}
\usepackage{bbm}
\usepackage{slashbox}
\usepackage{xcolor}
\usepackage{babel}

\def\bfgamma{\mbox{\boldmath $\gamma$}}

\def\bfsigma{\mbox{\boldmath $\sigma$}}

\def\bfvarepsilon{\mbox{\boldmath $\varepsilon$}}

\makeatletter
\@ifundefined{textcolor}{}
{%
 \definecolor{BLACK}{gray}{0}
 \definecolor{WHITE}{gray}{1}
 \definecolor{RED}{rgb}{1,0,0}
 \definecolor{GREEN}{rgb}{0,1,0}
 \definecolor{BLUE}{rgb}{0,0,1}
 \definecolor{CYAN}{cmyk}{1,0,0,0}
 \definecolor{MAGENTA}{cmyk}{0,1,0,0}
 \definecolor{YELLOW}{cmyk}{0,0,1,0}
}

\makeatother

\usepackage{babel}

\begin{document}

\title{Quasi Distribution Amplitude of Heavy Quarkonia}

\author{Yu Jia\footnote{jiay@ihep.ac.cn}}
\affiliation{Institute of High Energy Physics and Theoretical Physics Center for
Science Facilities, Chinese Academy of
Sciences, Beijing 100049, China\vspace{0.2cm}}
\affiliation{Center
for High Energy Physics, Peking University, Beijing 100871,
China\vspace{0.2cm}}

\author{Xiaonu Xiong\footnote{xiaonu.xiong@pv.infn.it}}

\affiliation{Istituto Nazionale di Fisica Nucleare, Sezione di Pavia, Pavia, 27100,
Italy}

\date{\today}

\begin{abstract}
The recently-proposed quasi distributions point out a promising direction
for lattice QCD to investigate the light-cone correlators,
such as parton distribution functions (PDF) and distribution amplitudes (DA),
directly in the $x$-space.
Owing to its excessive simplicity,
the heavy quarkonium can serve as an ideal theoretical laboratory to ascertain
certain features of quasi-DA.
In the framework of non-relativistic QCD (NRQCD) factorization, we compute the order-$\alpha_s$ correction to
both light-cone distribution amplitudes (LCDA) and quasi-DA associated with the lowest-lying quarkonia,
with the transverse momentum UV cutoff interpreted as the renormalization scale.
We confirm analytically that the quasi-DA of a quarkonium does reduce to
the respective LCDA in the infinite-momentum limit.
We also observe that, provided that the momentum of a charmonium reaches about
2-3 times its mass, the quasi-DAs already converge to the LCDAs to a decent level.
These results might provide some useful guidance for the future
lattice study of the quasi distributions.
\end{abstract}



\maketitle

\section{introduction}

The QCD factorization theorems~\cite{Collins:1989gx} imply that the
parton distribution functions (PDF)~\cite{Collins:1981uw} play the central role in accounting for virtually every
high-energy collision experiment.
In addition to PDF, there also exist other important types of light-cone correlators,
such as generalized parton distributions (GPD), transverse momentum dependent distributions
(TMDs), and light-cone distribution amplitudes (LCDA), all of which probe the internal structure of a hadron
in terms of fundamental quark-gluon degree of freedom.

These light-cone correlators are of nonperturbative nature, and are notoriously difficult to compute from
the first principle of QCD. The eminent obstacle for the lattice simulation originates from the fact that
they are defined in terms of the bilocal operators with light-like separation.
In the past,
lattice simulation has mainly focused on computing their
moments~\cite{Alexandrou:2015qia,Hagler:2007xi,Musch:2011er,Braun:2015lfa},
which are constructed out of the local operators.
Unfortunately, it becomes quickly impractical to go beyond a first few moments, since the more derivatives added,
the noisier the lattice simulation would become. To date,
our comprehensive knowledge about the nucleon PDF is gleaned exclusively through extracting from the
experimental data~\cite{Gao:2013xoa,Martin:2009iq,Ball:2014uwa}.

An exciting breakthrough has emerged recently.
A lattice calculation scheme directly in $x$-space was proposed by Ji in 2013~\cite{Ji:2013dva}.
In this approach, the task of computing the original light-cone correlators is transformed into computing
a new class of nonlocal matrix elements: the so-called {\it quasi} distributions.
These quasi distributions are defined as equal-time yet spatially-nonlocal correlation functions, thus amenable to
the lattice simulation.
In contrast to the light-cone quantities, the quasi distributions are generally frame-dependent.
But in the infinite momentum frame (IMF), the quasi distributions are expected to exactly recover the original
light-cone distributions. Ji has further envisaged that,
in analogy with the heavy quark effective theory (HQET), the quasi distribution method can be framed in an effective field theory context,
dubbed {\it Large Momentum Effective field Theory} (LaMET)~\cite{Ji:2014gla}.
The LaMET was first applied to proton spin structure,
which provides a means to extract the nucleon spin contents from the quasi distributions calculated on lattice~\cite{Ji:2013fga,Ji:2014lra}.

The utility of this new approach hinges crucially on the key that the quasi-distributions and
light-cone distribution share the exactly same infrared (IR) properties. It implies there exists a factorization theorem that connects these two quantities,
with perturbatively calculable matching coefficients.
Once the lattice has measured the quasi distributions, one can use this factorization formula to reconstruct
the desired light-cone quantities.

During the past two years, the one-loop matching factors have been computed for PDFs, GPDs for the non-singlet quark,
as well as pion DA~\cite{Xiong:2013bka,Ma:2014jla,Ji:2015qla}. The quasi TMD was also studied in \cite{Ji:2014hxa}.
Very recently, the two-loop renormalization of quasi-PDF has also been conducted~\cite{Ji:2015jwa}.
The factorization theorem for PDF has recently been proved to all orders in $\alpha_s$~\cite{Ma:2014jla}.
In addition, there recently have emerged some preliminary results from exploratory lattice simulations, extracting
the PDF from quasi PDF through the matching procedure outlined above~\cite{Lin:2014zya,Alexandrou:2015rja}.

To turn the quasi-distributions into a fruitful industry,
there remain many technical obstacles to overcome.
One outstanding challenge is to systematically implement the renormalization of such nonlocal operators on lattice.
Another difficulty stems from the technical limitation that, it is too expensive for the current lattice resources to
accommodate a fast-moving hadron on the lattice, since it requires exquisitely fine lattice spacing.
It is fair to say that, there is still a long way to go for the lattice simulation to be able to produce
phenomenologically competitive results.

For the lack of nonperturbative understanding of quasi distributions,
it is worth looking at their features from the perspective of phenomenological models.
For example, very recently the nucleon quasi-PDF has been investigated in a diquark model~\cite{Gamberg:2014zwa},
and the authors have examined how fast the nucleon quasi-PDF would approach the PDF with the increasing nucleon momentum.

Needless to say, it is also highly desirable to gain understanding about the gross features of the quasi distributions
from a model-independent angle. This consideration has motivated us
to study the distribution amplitudes (DA) of heavy quarkonia, chiefly because they offer
a unique, clean platform to scrutinize the quasi distributions. The key reason is that the DA of quarkonium
can be largely understood solely within perturbation theory.

The widely-separated scales ($m\gg mv,\,\Lambda_{\rm QCD}$) inherent to quarkonium
invites an effective-field-theory treatment.
In fact, the influential non-relativistic QCD (NRQCD) factorization approach~\cite{Bodwin:1994jh},
which fully exploits this scale hierarchy, nowadays has become an indispensable tool to tackle
quarkonium-related phenomena.

According to NRQCD factorization, the LCDA of a heavy quarkonium can be factorized as
the sum of the product of perturbatively-calculable, IR-finite coefficient functions and nonperturbative local NRQCD matrix elements~\cite{Ma:2006hc,Bell:2008er,Wang:2013ywc}.
At the lowest order in velocity expansion, up to a normalization factor,
the profile of the quarkonium LCDA is fully amenable to perturbation theory.

In this work, we generalize this knowledge and apply NRQCD factorization further to
the quasi-DA of heavy quarkonia, and calculate the respective coefficient functions to order $\alpha_s$.
To keep things as simple as possible, we concentrate on the lowest-lying $S$-wave quarkonia.
We have verified that, like the LCDAs of quarkonium, the quasi-DAs at order $\alpha_s$ are also IR-finite.
We are able to show analytically that, the quasi-DA exactly reduces to the LCDA in the infinity-momentum limit.
We also observe that, provided that the quarkonium is boosted to carry a momentum about 2-3 times its mass,
and with the renormalization scale chosen around the charmonium mass,
the respective quasi-DAs will converge to the LCDAs to a satisfactory degree.

We hope some of features about the quarkonium quasi-DAs may also apply to other hadrons.
Hopefully this knowledge will provide some useful guidance to the future lattice investigation
of similar quasi distributions.

The rest of the paper is organized as follows.
In Sec.~\ref{NRQCD:fac:LC:quasi}, we present the definitions of LCDA and quasi-DA for $S$-wave quarkonia,
and discuss the precise meaning of NRQCD factorization to these correlators.
In Sec.~\ref{LCDA:quasi-DA:one:loop:results}, we describe the strategy to determine the DAs of quarkonia
beyond tree-level, and outline the key steps of deriving the one-loop corrections,
and present the corresponding analytical expressions for the $S$-wave quarkonia.
In Sec.~\ref{Numerical:study:DA}, we carry out numerical comparison between LCDA and quasi-DA,
to study how fast the quasi-DA approaches the LCDA as the quarkonium momentum increases.
We also compare the first inverse moments calculated in both LCDA and quasi-DA.
We summarize in Sec.~\ref{summary}.
The detailed illustrations about how to work out the one-loop calculation are provided
in the Appendices.

\section{NRQCD factorization of Quarkonium Distribution Amplitudes}
\label{NRQCD:fac:LC:quasi}

In contrast to the light hadrons, heavy quarkonia are arguably among the simplest hadrons:
its constituent quark and antiquark are quite heavy, $m\gg \Lambda_{\rm QCD}$,
and move rather slowly ($v\ll 1$). These two essential features result in the
hierarchical structure of intrinsic energy scales of a quarkonium.
NRQCD factorization approach~\cite{Bodwin:1994jh} fully exploits this scale hierarchy,
and allows one to efficiently separate the relativistic/perturbative contributions from the long-distance/nonperturbative dynamics.
For most quarkonium-related phenomena, {\it i.e.} quarknoium production and decay processes,
this factorization approach has become an standard tool.

It is well known that the fragmentation functions for a parton transitioning into a light hadron
are genuinely nonperturbative
objects, and the only way to extract them is through experimental measurements~\cite{Agashe:2014kda}.
On the contrary, it was realized long ago that the heavy quarkonium fragmentation function can be
put in a factorized form~\cite{Braaten:1993mp,Braaten:1993rw}.
Concretely speaking, for a gluon-to-quarkonium fragmentation function, one has
\begin{align}
D_{g\to H+X}(z,\mu) &= \sum_n  d_{g\to c\bar{c}[n]}(z) \langle 0| O_H[n] |0\rangle,
\end{align}
where $z$ denotes the momentum fraction, and $n$ specifies the color/spin/orbital quantum number of the $c\bar{c}$ pair, and
$O_H[n]$ is the NRQCD four-fermion operators, which characterizes the transition probability from the
partonic state $c\bar{c}[n]$ to the quarkonium $H$ plus additional soft hadrons.
The key insight is that coefficient functions $d_{g\to c\bar{c}[n]}(z)$ are perturbatively calculable.

Analogous to the case of aforementioned fragmentation function,
one might naturally envisage that the DA of a quarkonium is also not a fully nonpertubative object,
and some sort of short-distance ($\sim 1/m$) effects should be disentangled owing to asymptotic freedom.
Indeed, such an analogy has already been pursued some time ago~\cite{Ma:2006hc,Bell:2008er}.
Schematically, one may express the quarkonium DA in the following factorized form:
\begin{align}
\Phi_{H}(x,\mu) \sim & \sum_{n} \langle H\left| \mathcal{O}_{[n]}\right|0\rangle
\phi_{H[n]}(x,\mu),
\label{eq:ReFact}
\end{align}
where the color-singlet NRQCD operators $\mathcal{O}_{[n]}$ are
organized according to the importance in the velocity expansion. Apart from the universal NRQCD matrix elements,
the key observation is that $\phi_{H[n]}\left(x,\mu\right)$ can now be interpreted as the short-disance coefficients.
Actually, for the hard exclusive quarkonium production, employing this factorized quarkonium LCDA turns out to
have considerable advantage compared with conventional NRQCD factorization approach~\cite{Jia:2008ep,Jia:2010fw}.

For simplicity, in this work we will only concentrate on the distribution amplitudes of $S$-wave quarkonia. Moreover,
we will only be interested in the lowest order in $v$ expansion.
Obviously, there is no any principal difficulty to incorporate the relativistic corrections,
or even extend to higher orbital quarkonium states.

\subsection{NRQCD factorization of quarkonium LCDA}

To be specific, let us assume the quarkonium $H$ to move along the positive $z$-axis,
{\it i.e.}, $P^{\mu}=\left(\sqrt{P_z^2+m^{2}},\boldsymbol{0}_{\perp}, P^{z} \right)$ with $P^z>0$.
For a general 4-vector $V^\mu$, it is convenient to introduce the light-cone plus (minus) components
 $V^{\pm}={1\over \sqrt{2}}(V^{0}\pm V^{z})$.

The leading-twist LCDAs of the pseudoscalar meson $P$, longitudinally (transversely) polarized vector meson $V^{\parallel,\perp}$,
are defined as
\begin{subequations}
\begin{align}
\Phi_{P}\left(x,\mu\right)= & -if_{P} P^{+} \phi_{P}\left(x,\mu\right)\nonumber \\
= & \int\frac{d\xi^{-}}{2\pi}\,e^{-i\left(x-\frac{1}{2}\right) P^{+}\xi^{-}}\left\langle P\left(P\right)\!\left|\!\bar{\psi}\left(\frac{\xi^{-}}{2}\right)\gamma^{+}\gamma^{5}\mathcal{W}
\psi\left(-\frac{\xi^{-}}{2}\right)\!\right|0\right\rangle ,\\
\Phi_{V}^{\parallel}\left(x,\mu\right)= & -if_{V}^{\parallel}\varepsilon_{\parallel}^{*+} M_{V} \phi_{V}^{\parallel}\left(x,\mu\right)\nonumber \\
= & \int\frac{d\xi^{-}}{2\pi}\,e^{-i\left(x-\frac{1}{2}\right) P^{+}\xi^{-}}\left\langle V\left(P,\varepsilon_\parallel\right)\!\left|\!\bar{\psi}\left(\frac{\xi^{-}}{2}\right)\gamma^{+}
\mathcal{W} \psi\left(-\frac{\xi^{-}}{2}\right)\!\right|0\right\rangle ,\\
\Phi_{V}^\perp\left(x,\mu\right)= & -i f_{V}^{\perp} P^{+} \phi_{V}^\perp \left(x,\mu\right)
\nonumber \\
= & \int\frac{d\xi^{-}}{2\pi}\,e^{-i\left(x-\frac{1}{2}\right)P^{+}\xi^{-}}\left\langle V\left(P,\varepsilon_\perp \right)\!\left|\!\bar{\psi}\left(\frac{\xi^{-}}{2}\right)\gamma^{+} \bfgamma \cdot \bfvarepsilon_{\perp}
\mathcal{W} \psi\left(-\frac{\xi^{-}}{2}\right)\!\right|0\right\rangle,
\end{align}
\label{eq:LCDA_Def}
\end{subequations}
where $\varepsilon_{\parallel}^{\mu}$,
$\epsilon_{\perp}^{\mu}$ are the polarization vector for longitudinally
and transversely polarized vector meson, $\mu$ signifies the renormalization scale. $\mathcal{W}$ is the gauge
link along the light-cone ``minus'' direction:
\begin{align}
\mathcal{W}  &= {\mathcal P} \exp\left[-i g_s
\int^{\xi^{-}\over 2}_{-{\xi^{-}\over 2}}
d \eta^- A^+(\eta^-)\right].
\end{align}

The decay constants $f_{H}$ are defined as the vacuum-to-quarkonium
matrix elements mediated by various local QCD currents:
\begin{subequations}
\begin{align}
\langle P (P) |\bar{\psi}\gamma^{+}\gamma^{5}\psi|0\rangle & \equiv  -i  f_{P} P^{+}= \int_{0}^{1}dx\,\Phi_{P}\left(x,\mu\right),\\
\left\langle V(P,\varepsilon_{\parallel})\left|\bar{\psi} \gamma^{+}\psi \right|0\right\rangle & \equiv -i M_V
f_V^{\parallel} \varepsilon_{\parallel}^{*+} = \int_{0}^{1}dx\,\Phi_{V}^{\parallel} \left(x,\mu\right),\\
\left\langle V(P,\varepsilon_{\perp})\left| \bar{\psi} \gamma^+ \bfgamma_\perp \psi \right|0\right\rangle & \equiv
-i f_{V}^\perp P^{+} \bfvarepsilon_{\perp}^{*}= \int_{0}^{1}dx\,\Phi_V^\perp(x,\mu).
\end{align}
\label{Def:decay:constant:LC}
\end{subequations}

The LCDA is clearly subject to the normalization condition:
\begin{align}
\int^1_0 \! dx\,\phi_H(x)=1 \qquad{\rm for} \;\;\forall\;\; H.
\end{align}

Thus far, everything is about the standard definition, valid for any pseudoscalar and vector mesons.
So what is special about the heavy quarkonium? As has been argued previously, the quarkonium DA defined above
still contains short-distance contribution, which ought to be identified and isolated.

If $H$ is a $S$-wave quarkonium state, the precise implication of NRQCD factorization of the LCDA is
\begin{subequations}
\begin{align}
\phi_H(x) &= \phi^{(0)}_H(x) + {C_F\alpha_s  \over \pi}\phi^{(1)}_H(x) +\cdots, \label{LCDA:pert:expansion} \\
f_{H} &= f_{H}^{(0)} \left(1+{C_F\alpha_s \over \pi}\,{\mathfrak f}_{H}^{(1)}+\cdots \right)+O(v^2), \label{Decay:constant:matching}
\end{align}
\label{Precise:meaning:NRQCD;fac:LCDA}
\end{subequations}
where $H= P, V_\parallel, V_\perp$. For the DAs of the hidden-flavor quakonia (the $c\bar{c}$ or $b\bar{b}$ family),
charge conjugation symmetry demands that they are symmetric under $x\leftrightarrow 1-x$.

The key message conveyed in (\ref{Precise:meaning:NRQCD;fac:LCDA}) is that the
$\phi_H(x)$ entailing all the hard ``collinear'' degree of freedom
(with typical virtuality of order $m^2$), thus can be computed in perturbation theory owing to asymptotic freedom.
The nonperturbative aspects of quarkonium are encoded
in the decay constant $f_H$.
Moreover, as indicated in (\ref{Decay:constant:matching}), one can match the QCD currents to the respective NRQCD
currents, by integrating out the hard quantum fluctuation. Consequently,
the genuinely nonperturbative binding dynamics is
encapsulated in the NRQCD matrix elements $f_{H}^{(0)}$.
For $H=\eta_c,\,J/\psi$, one has
\begin{subequations}
\begin{align}
f_{\eta_c}^{(0)} &= {1\over \sqrt{m_c}} \langle \eta_c| \psi^\dagger \chi|0\rangle  \approx
\sqrt{N_c\over 2\pi m_c} R_{\eta_c}(0), \\
f_{J/\psi}^{\parallel(0)} &= f_{J/\psi}^{\perp (0)}= {1\over \sqrt{m_c}} \langle J/\psi(\bfvarepsilon)| \psi^\dagger\bfsigma\cdot
\bfvarepsilon \chi|0\rangle \approx \sqrt{N_c\over 2\pi m_c} R_{J/\psi}(0),
\end{align}
\label{Def:NRQCD:vac:to:H:matrix:element}
\end{subequations}
where $\bfvarepsilon$ denotes the polarization three-vector in the $J/\psi$ rest frame,
and $N_c=3$ is the number of colors in QCD. Since NRQCD matrix elements are always defined in the quarkonium rest frame,
rotation invariance then implies that $f_{J/\psi}^{\parallel(0)} = f_{J/\psi}^{\perp(0)}$.
As implied in the last entity, these NRQCD matrix elements are often approximated by $R_H(0)$,
the radial Schr\"{o}dinger wave function at the origin for the $S$-wave charmonia,
which can be evaluated in the phenomenological quark potential models.

\subsection{NRQCD factorization of quarkonium quasi DA}

The quasi DAs are defined as pure spatial correlation
functions, hence can be directly simulated on the lattice.
Analogous to LCDA (\ref{eq:LCDA_Def}), we define the quasi-DAs of $S$-wave quarkonia as
\begin{subequations}
\begin{align}
\widetilde{\Phi}_{\eta_{c}}\left(x,\mu,P^{z}\right)= & -i\tilde{f}_{\eta_{c}}P^{z}\tilde{\phi}_{\eta_{c}}\left(x,\mu\right)\nonumber \\
= & \int\frac{dz}{2\pi}\,e^{i\left(x-\frac{1}{2}\right)P^{z}z}\left\langle \eta_{c}\left(P\right)\!\left|\!\bar{\psi}\left(\frac{z}{2}\right)\gamma^{z}\gamma^{5}
\mathcal{V} \psi\left(-\frac{z}{2}\right)\!\right|0\right\rangle ,\\
\widetilde{\Phi}_{\eta_{c}}^{\parallel}\left(x,\mu,P^{z}\right)=
& -i\tilde{f}_{J/\psi}^{\parallel}\varepsilon_{\parallel}^{*z} M_{J/\psi} \tilde{\phi}_{J/\psi}^{\parallel} \left(x,\mu\right)
\nonumber \\
= & \int\frac{dz}{2\pi}\,e^{i\left(x-\frac{1}{2}\right)P^{z}z}\left\langle J/\psi \left(P, \varepsilon_\parallel \right)\!\left|\!\bar{\psi}\left(\frac{z}{2}\right)\gamma^{z}
\mathcal{V} \psi\left(-\frac{z}{2}\right)\!\right|0\right\rangle,\\
\widetilde{\Phi}_{J/\psi}^{\perp}\left(x,\mu,P^{z}\right)=
& -i\tilde{f}_{J/\psi}^{\perp} P^{z}\phi_{J/\psi}^{\perp} \left(x,\mu\right)\nonumber \\
= & \int\frac{dz}{2\pi}\,e^{i\left(x-\frac{1}{2}\right) P^{z} z}\left\langle J/\psi\left(P,\varepsilon_\perp\right)\!\left|\!\bar{\psi}\left(\frac{z}{2}\right)\gamma^{z}\boldsymbol{\gamma}\cdot \boldsymbol{\varepsilon}_{\perp}^{*}
\mathcal{V} \psi
\left(-\frac{z}{2}\right)\!\right|0\right\rangle,
\end{align}
\label{eq:quaDA_Def}
\end{subequations}
where the field separation is along
the $z$ direction, and the gauge link
$\mathcal{V}$ reads
\begin{align}
\mathcal{V}  &= {\mathcal P} \exp\left[-i g_s \int^{z\over 2}_{-{z\over 2}}
d \eta^z A^z(\eta^z)\right].
\end{align}

The quasi decay constants, dubbed $\tilde{f}_H$, are again defined as the vacuum-to-quarkonium
matrix elements mediated by corresponding QCD currents~\footnote{Lorentz invariance requires $\tilde{f}_H=f_H$.
Here we intentionally distinguish these two cases, because one may choose a UV regulator
that does not preserve Lorentz symmetry.
In the loop integrals, we will impose a UV cutoff in the transverse momentum components,
which does not to violate the boost invariance along $z$ axis. Therefore, in our case,
we indeed have $\tilde{f}_H=f_H$, and will use them interchangeably.}:
\begin{subequations}
\begin{align}
\langle \eta_c (P) |\bar{\psi}\gamma^{z}\gamma^{5}\psi|0\rangle & \equiv  -i  \tilde{f}_{P} P^{z}= \int_{-\infty}^{\infty} \! dx\,\widetilde{\Phi}_{P}\left(x,\mu\right),\\
\left\langle J/\psi(P,\varepsilon_{\parallel})\left|\bar{\psi} \gamma^{z}
\psi \right|0 \right\rangle
& \equiv -i M_{J/\psi} \tilde{f}_{J/\psi}^{\parallel} \varepsilon_{\parallel}^{*z} =
\int_{-\infty}^{\infty}\! dx\,\widetilde{\Phi}_{J/\psi}^{\parallel} \left(x,\mu\right),\\
\left\langle J/\psi(P,\varepsilon_{\perp})\left| \bar{\psi} \gamma^z \bfgamma_\perp \psi \right|0\right\rangle & \equiv
-i f_{J/\psi}^\perp P^{z} \bfvarepsilon_{\perp}^{*} = \int_{-\infty}^{\infty} \! dx\, \widetilde{\Phi}_{J/\psi}^\perp(x,\mu).
\end{align}
\end{subequations}

The NRQCD factorization is also valid for the quasi-DAs.
For the quasi DA of a $S$-wave quarkonia,
the precise implication of NRQCD factorization is
\begin{subequations}
\begin{align}
\tilde{\phi}_H(x) &= \tilde{\phi}^{(0)}_H(x) + {C_F\alpha_s(\mu) \over \pi} \,\tilde{\phi}^{(1)}_H(x) +\cdots,\\
\tilde{f}_{H} &= f_{H}^{(0)} \left(1+{C_F\alpha_s(2m_c) \over \pi}\,{\mathfrak f}_{H}^{(1)}+\cdots \right)+O(v^2), \label{quasi:Decay:constant:matching}
\end{align}
\end{subequations}
where $H= P, V_\parallel, V_\perp$.
The matching of the decay constant is exactly the same as in (\ref{Decay:constant:matching}).

The quasi-DA is subject to the normalization:
\begin{align}
\int^\infty_{-\infty} \tilde{\phi}_H(x)=1 \qquad{\rm for} \;\;\forall\;\; H.
\end{align}
Because quasi-DA no longer emerges from a parton picture,
the integrand is no longer bounded within the interval $x\in[0,1]$.

In contrast to the LCDA, the quasi-DA is generally dependent on the magnitude of $P^z$.
We note that the heavy quark mass in NRQCD factorization corresponds to a large scale,
so it cannot be neglected even in the limit $P^z\gg m$.

\section{One-loop expressions of distribution amplitudes for $S$-wave quarkonia}
\label{LCDA:quasi-DA:one:loop:results}

In this section, we compute the one-loop corrections to both LCDAs and quasi-DAs for the
$S$-wave quarkonia, to lowest order in $v$.

\subsection{Strategy of determining the LCDA and quasi-DA for quarkonium}

The $\phi_H(x)$ ($\tilde{\phi}_H(x)$) is only sensitive to short-distance dynamics.
In order to extract it, it is convenient to replace a physical quarkonium state $|H(P)\rangle$ by a fictitious one,
{\it i.e.}, a pair of free heavy quark-antiquark $|c(p_1)\bar{c}(p_2)\rangle$.
We then compute the corresponding $\Phi_{c\bar{c}(P)}(x)$ in perturbation theory:
\begin{subequations}
\begin{align}
\Phi_{c\bar{c}(P)}(x) &= \Phi_{c\bar{c}}^{(0)}(x) + {C_F\alpha_s(\mu) \over \pi} \Phi_{c\bar{c}}^{(1)}(x) +\cdots, \\
\widetilde{\Phi}_{c\bar{c}(P)}(x) &= \widetilde{\Phi}_{c\bar{c}}^{(0)}(x) + {C_F\alpha_s(\mu) \over \pi} \widetilde{\Phi}_{c\bar{c}}^{(1)}(x) +\cdots.
\end{align}
\end{subequations}
The partonic decay constant $f_{c\bar{c}(P)}$ can also be computed order by order in $\alpha_s$.
Following the definitions in (\ref{eq:LCDA_Def}) and (\ref{eq:quaDA_Def}),
after projecting onto the suitable quantum number, one should be able
to solve for $\phi_H(x)$ ($\tilde{\phi}_H(x)$) iteratively, order by order in $\alpha_s$.

The $c$ and $\bar{c}$ in the fictitious charmonium state $|c(p_1)\bar{c}(p_2)\rangle$
carry momenta $p_1={P\over 2}+q$ and $p_2={P\over 2}-q$, respectively.
Since we are only interested in the lowest order in $v$, thus it is legitimate to neglect the relative momentum
$q$, from now on we thereby assume $p_1=p_2={P\over 2}\equiv p$.

When going beyond the tree level, rather than utilize the literal matching method, we take a standard shortcut to directly extracting the short-distance coefficient (arising from the hard region $~m^2$) in the loop integral~\cite{Beneke:1997zp}: in the beginning, we simply neglect the relative momentum $q$ prior to carrying out the loop integration~\footnote{For a one-loop computation of the $S$-wave quarkonium LCDA following the rigorous NRQCD matching ansatz, we refer the interested readers to Ref.~\cite{Ma:2006hc}.}.
Therefore, our calculation is free from the contamination due to the low-energy effects
(loop momentum carrying virtuality of order $mv$ or smaller, exemplified by the Coulomb singularity).
This brings forth great technical simplification.
Nevertheless, the general principle of effective field theory guarantees that the contributions from the low-energy
regimes must cancel between the QCD side and NRQCD side,
and we simply trust it holds and forgo this check.

\subsection{Tree-level results}

At the lowest order in $\alpha_s$, it is straightforward to work out the
partonic DAs:
\begin{subequations}
\begin{align}
\Phi^{(0)}_{c(p)\bar{c}(p)}\left(x,\mu\right)
& = \int\frac{d\xi^{-}}{2\pi}\,e^{-i\left(x-\frac{1}{2}\right) P^{+}\xi^{-}}\left\langle c(p)\bar{c}(p)\!\left|\!\bar{\psi}\left(\frac{\xi^{-}}{2}\right)\gamma^{+}\Gamma
\psi\left(-\frac{\xi^{-}}{2}\right)\!\right|0\right\rangle\nonumber \\
&= \delta\left(x-\frac{1}{2}\right) {1\over P^+}\bar{u}(p)\gamma^{+}\Gamma v(p),
\\
\widetilde{\Phi}^{(0)}_{c(p)\bar{c}(p)}\left(x,\mu,P^{z}\right) &= \int\frac{dz}{2\pi}\,e^{i\left(x-\frac{1}{2}\right)P^{z}z}\left\langle c(p)\bar{c}(p)\!\left|\!\bar{\psi}\left(\frac{z}{2}\right)\gamma^{z}\Gamma
\psi\left(-\frac{z}{2}\right)\!\right|0\right\rangle
\nonumber \\
&= \delta\left(x-\frac{1}{2}\right) {1\over P^z}\bar{u}(p)\gamma^{z}\Gamma v(p),\\
\langle c(p)\bar{c}(p) |\bar{\psi}\not\! n \Gamma \psi|0\rangle & = {1\over n\cdot P}\bar{u}(p)\not\! n \Gamma v(p),
\end{align}
\end{subequations}
where we have introduced the reference vector $n$:
\begin{align}
n^{\mu} = & \begin{cases}
 {1\over \sqrt{2}}(1,0,0,-1)\qquad\qquad{\rm for\;LCDA}, \\
 (0,0,0,-1)\:\quad\qquad\qquad{\rm for\;quasi\!\!-\!\!DA}.
\end{cases}
\label{reference:vector:n:def}
\end{align}
$\Gamma=\gamma^5,\boldsymbol{1},\gamma^i_\perp$ correspond to $\eta_c$, longitudinally and transversely polarized $J/\psi$ meson, respectively.

Since we work with a moving frame of $H$, it is convenient to adopt the threshold expansion method developed by Braaten and Chen~\cite{Braaten:1996jt}, which takes into account of the Lorentz transformation between quarkonium rest frame and the moving frame, making the connection to NRQCD transparent. One then finds
\begin{subequations}
\begin{align}
{1\over n\cdot P}\bar{u}(p) \not\!n \gamma_5 v(p) &= \xi^\dagger \eta \bigg|_{c\bar{c}\;{\rm rest\;\,frame}},\\
{1\over P^0}\bar{u}(p)\gamma^{z} v(p) &= \xi^\dagger \sigma^z\eta \bigg|_{c\bar{c}\;{\rm rest\;\,frame}}, \\
{1\over P^+}\bar{u}(p)\gamma^{+} v(p) &= \xi^\dagger \sigma^z\eta \bigg|_{c\bar{c}\;{\rm rest\;\,frame}}, \\
{1\over n\cdot P}\bar{u}(p) \not\!n \gamma_\perp^i v(p) &= \xi^\dagger \sigma_\perp^i\eta \bigg|_{c\bar{c}\;{\rm rest\;\,frame}},
\end{align}
\end{subequations}
where $\xi$ and $\eta$ are two-components Pauli spinors.
When the $c\bar{c}$ pair is in the $^3S_1(\bfvarepsilon)$ state,
$\xi^\dagger \bfsigma\eta \propto \bfvarepsilon$.
Everything has  the desired structure as dictated in (\ref{Def:decay:constant:LC}), especially the Lorentz transformation of longitudinal polarization vector is correctly incorporated.
From these knowledge, we can readily determine $f_{c\bar{c}}^0$
for $c\bar{c}({}^1S_0)$ and $c\bar{c}({}^3S_1)$.

Therefore, the tree-level LCDAs and quasi-DAs bear the simple form:
 \begin{align}
  \phi^{(0)}_H(x) &=  \tilde{\phi}^{(0)}_H(x) = \delta\left(x-{1\over 2}\right).
\end{align}
where $H= P, V_\parallel, V_\perp$.
Obviously it satisfies the normalization condition (\ref{Def:decay:constant:LC}).

Intuitively, this is what is expected from the nonrelativistic limit,
when the relative momentum is neglected.


\subsection{Outline of one-loop calculation}

The DAs of quarkonium will develop nontrivial profile after implementing radiative correction.
Its shape generally becomes widely spread.
This extended profile should not be confused with the LCDAs determined by phenomenoglical models such as QCD sum rules~\cite{Braguta:2006wr},
because it is generated perturbatively and can be computed in a model-independent manner.

We now turn to calculating the order-$\alpha_s$ correction to LCDA and quasi-DA, that is, to determine $\phi_H^{(1)}$ and $\tilde{\phi}_H^{(1)}$. This can be fulfilled by employing the following relation:
\begin{subequations}
\begin{align}
{\phi}_H^{(1)}(x) &= \Phi_{c\bar{c}}^{(1)}\left(x\right)\bigg/\left({1\over P^+}\bar{u}(p)\gamma^{+}\Gamma v(p)\right)-
\delta\left(x-{1\over 2}\right)\, \mathfrak{f}^{(1)},
\\
 \tilde{\phi}_H^{(1)}(x) &= \tilde{\Phi}_{c\bar{c}}^{(1)}\left(x\right)\bigg/\left({1\over P^z}\bar{u}(p)\gamma^{z}\Gamma v(p)\right)- \delta\left(x-{1\over 2}\right)\, \mathfrak{f}^{(1)}.
\end{align}
\label{strategy:deter:one:loop:DA}
\end{subequations}
Therefore, we need calculate the order-$\alpha_s$ correction to partonic (quasi-)DA and corresponding decay constant.

\begin{figure}[tbH]
\centering{}\includegraphics[scale=0.75]{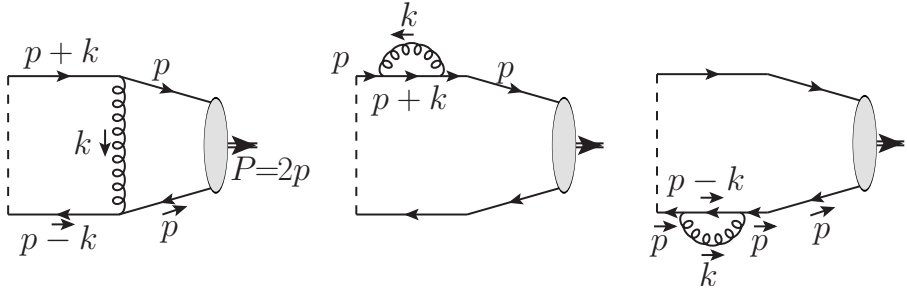}
\protect\caption{One-loop diagrams for the S-wave quarkonium (quasi-)DA in axial gauge.}
\label{DA_1loopFeyn}
\end{figure}

Although LCDA and quasi-DA by construction are gauge invariant objects,
practically we have to specify a gauge when computing the one-loop correction.
We find it convenient to work with the axial gauges: {\it i.e.}, $A^+=0$ for LCDA, and
$A^z=0$ for quasi-DA.
In such gauges, the gauge link shrinks to unity, and we only have to deal with very few diagrams,
which are depicted in Fig.~\ref{DA_1loopFeyn}.
Now the complication instead resides in the gluon propagator:
\begin{equation}
D_{\mu\nu}\left(k\right)= \frac{-i}{k^{2}+i\epsilon}\left(g_{\mu\nu}-\frac{n_{\mu}k_{\nu}+n_{\nu}k_{\mu}}{n\cdot k}+\frac{n^{2}k_{\mu}k_{\nu}}{n\cdot k^{2}}\right).
\label{def:gluon:propagator:axial:gauge}
\end{equation}
where $n^\mu$ is defined in (\ref{reference:vector:n:def}).

Ultraviolet divergences will inevitably emerge in our calculation, thereby necessitating the introduction of a UV regulator.
For the light-cone correlators such as PDF and LCDA, only the logarithmic UV divergences will arise; nevertheless, for the
quasi distributions, one often confronts linear or even severer UV divergences~\cite{Xiong:2013bka,Ma:2014jla}.
In order to keep track of these violent UV divergences, it is more transparent to adopt a physical UV regulator such as a hard momentum
cutoff than simply use the dimensional regularization (DR).
In some sense, the UV cut-off $\Lambda$ imposed on the transverse-momentum integration may be viewed as intimately mimicking the role placed by the lattice spacing in lattice Monte Carlo simulation.
In this work, we will also utilize the transverse momentum cutoff $\Lambda$ to regularize the UV divergence.

Thus far, the renormalization program of nonlocal correlators, particularly the quasi distributions,
has not yet been fully developed, and remains as an active research topic~\cite{Ma:2014jla,Ji:2015jwa}.
The hope is that the UV divergences associated with the quasi distributions can be removed through the multiplicative renormalization to all orders in $\alpha_s$~\cite{Ma:2014jla,Ji:2015jwa}.
As a consequence, a rigorous renormalization procedure of the quasi-DA is beyond the scope of the current work.
Our primary goal in this work is to compare the behavior of quasi-DA and LCDA at variance with $P^z$.
For this purpose, $\Lambda$ will be kept finite and taken around the characteristic heavy quark mass scale.
Roughly speaking, we pretend to have a ``renormalized'' LCDA and quasi-DA with $\Lambda$
interpreted as the corresponding renormalization scale $\mu$ in a continuum quantum field theory.

Another practical reason for us to keep $\Lambda$ finite is because the order of taking two limits
$\Lambda\rightarrow\infty$ and $P^{z}\rightarrow\infty$ is not commutable. Had $\Lambda\rightarrow\infty$
been taken first, the quasi distributions would not approach the light-cone
distributions even in the limit $P^{z}\rightarrow\infty$. As the main goal of
this paper is to investigate  quantitatively how the quasi-DA can approach the LCDA with increasing $P^z$,
therefore, the analytic control of $P^{z}\rightarrow\infty$ limit is a crucial requirement.
For this purpose, keeping a finite $\Lambda$ is crucial.

It is worth pointing out that, besides UV divergences, IR divergences also arise from individual diagrams in Fig.~\ref{DA_1loopFeyn}.
It can be traced from the exchange of soft gluon between quark and antiquark that equally partition the
total momentum $P$, so are always accompanied with $\delta(x-{1\over 2})$. Of course, when summing the vertex diagram and the quark self-energy diagram together, the IR singularities cancels, as ensured by the validity of NRQCD factorization:
the color-singlet NRQCD bilinears such as $\psi^\dagger \chi$ and $\psi^\dagger \bfsigma \chi$
do not acquire an anomalous dimension at order-$\alpha_s$~\footnote{
If we attempt to extract the two-loop correction to (quasi-)DAs using the same technique as described in this work,
we would confront the uncancelled single IR pole, which should be absorbed into the corresponding vacuum-to-quarkonium NRQCD matrix elements. It is intimately linked to the fact that NRQCD quark bilinears $\psi^\dagger \chi$ and $\psi^\dagger \bfsigma \chi$
acquire an anomalous dimension first at order-$\alpha_s^2$~\cite{Czarnecki:1997vz,Beneke:1997jm,Czarnecki:2001zc}.}.
However, it is still necessary to introduce an IR regulator in the intermediate steps.
In our calculation, we find it convenient to employ the DR to regularize the IR singularity, working with
the spacetime dimension $d=4-2\epsilon$ ($\epsilon<0$)~\footnote{In previous calculation of
the LCDA of the $S$-wave quarkonium using NRQCD factorization~\cite{Bell:2008er,Wang:2013ywc}, the authors employ the DR to
regularize UV and IR divergences simultaneously. If taking $\Lambda\to \infty$ prior to Laurent-expanding $\epsilon$, we would be able to reproduce their unrenormalized one-loop results.}.
We stress that the popular way of introducing a fictitious small gluon mass is not sufficient to tame the IR divergences encountered here, because in our calculation the soft IR singularity can be coupled with the
axial singularity (stemming from $1/n\cdot k$), where the latter can not be regularized by just adding $m_g$ alone~\footnote{Precisely speaking, the most singular IR behavior is captured by the ${1\over \epsilon^2_{\rm IR}}$ pole if DR is used. If one uses the gluon mass regularization for the soft divergence, one has to invoke additional regulator such as DR to regularize the axial singularity, so the severest IR singularity would look like ${1\over \epsilon_{\rm IR}}\ln m_g$ as $x\to {1\over 2}$.}.

For the one-loop integral, we always choose to first integrate over the $k^-$ ($k^0$) component
using contour technique for LCDA (quasi-DA),
then carry out the remaining $d-2$-dimensional integration over transverse components, finally end up with a
one-dimensional integral depending on the variable $k^+$ ($k^z$) for LCDA (quasi-DA). Then one can readily read off the desired distribution as a function of x, which is connected with $k^+$ ($k^z$) through the relation $k^+=(x-1/2)P^+$ ($k^z=(x-1/2)P^z$) for LCDA (quasi-DA) that is enforced by $\delta$ function.
The following transverse-momentum integration measure is ubiquitously encountered:
\begin{equation}
\left(\frac{\mu_{\rm IR}^{2}e^{\gamma_E}}{4\pi}\right)^{\epsilon}\int d^{2-2\epsilon}k_{\perp}=\left(\frac{\mu_{\rm IR}^{2}e^{\gamma_E}}{4\pi}\right)^{\epsilon}
\frac{2\pi^{1-\epsilon}}{\Gamma\left(1-\epsilon\right)}\int_{0}^{\Lambda}k_{\perp}^{1-2\epsilon}dk_{\perp},
\end{equation}
where $\Lambda$ is the UV cut-off, $\gamma_E$ is the Euler constant, and $\mu_{\rm IR}$ is the 't Hooft's unit mass.
We put an subscript ``IR'' to emphasize this scale is affiliated with the IR divergence.

For both LCDA and quasi-DA, the vertex diagram in Fig.~\ref{DA_1loopFeyn} can be written as
\begin{align}
\Phi_{c\bar{c}}^{(1)\,\rm ver}= & g_{s}^{2}C_{F}\left(\frac{\mu_{\rm IR}^{2} e^{\gamma_E}}{4\pi}\right)^{\epsilon}
\int\frac{d^{4-2\epsilon}k}{\left(2\pi\right)^{4-2\epsilon}}\,\bar{u}\left(p\right)
\gamma^{\mu}\frac{1}{k\!\!\!/+p\!\!\!/-m+i\epsilon}n\!\!\!/\left\{ \gamma^{5},\boldsymbol{1},\gamma^{\perp}\right\} \nonumber \\ & \times\frac{1}{k\!\!\!/\!-\!p\!\!\!/\!-\!m\!+\!i\epsilon}\gamma^\nu v\left(p\right)D_{\mu\nu}\left(k\right)\!\delta\!\left(\!\!x\!-\!\frac{1}{2}\!-\!\frac{n\cdot k}{n\cdot P}\!\!\right).
\label{eq:DA:one:loop:vertex}
\end{align}
The three terms associated with the gluon propagator have a one-to-one
correspondence with what would be encountered in a Feynman-gauge calculation.
The second and third terms correspond to the diagrams entailing a gauge-link
interaction, Particularly, the third term will correspond to a self-energy correction to the gauge link,
would actually lead to a linear UV divergence for quasi DA.

In Appendix~\ref{1loopExmpl}, we have provided comprehensive details on how to work out the
Feynman part ($\propto g_{\mu\nu}$ in the gluon propagator) of this one-loop vertex integral.
After accomplishing all the algebra, it is reassuring that $\Phi_{c\bar{c}}^{(1)\,\rm ver}$ turns out to possess the exactly same Lorentz structure as the tree-level format,
$\propto {1\over P^+}\bar{u}(p)\gamma^{+}\Gamma v(p)$.
This feature is in conformity with Eq.~(\ref{strategy:deter:one:loop:DA}).

According to the LSZ reduction formula, we also have to include
the order-$\alpha_s$ correction to the quark wave function renormalization constant.
It only yields a $\delta(x-{1\over 2})$ piece to $\Phi_{c\bar{c}}(x)$ ($\widetilde{\Phi}_{c\bar{c}}(x)$).
The contributions from last two diagrams in Fig.~\ref{DA_1loopFeyn} read
\begin{subequations}
\begin{align}
\Phi_{c\bar{c}}^{(1)\,\rm wvf} &= {1\over 2}\left[\delta Z_{F,q}^{(1)}+\delta Z_{F,\bar{q}}^{(1)}\right]\delta\left(x-\frac{1}{2}\right),
\\
\widetilde{\Phi}_{c\bar{c}}^{(1)\,\rm wvf} &= {1\over 2}\left[\delta\tilde{Z}_{F,q}^{(1)}+\delta\tilde{Z}_{F,\bar{q}}^{(1)}\right]\delta\left(x-\frac{1}{2}\right).
\end{align}
\end{subequations}
The quark wave function renormalization constant $Z_F$ in axial gauges are considerably more
complicated than its counterpart in covariant gauges.

We follow the recipe given in \cite{Xiong:2013bka} to express them as
\begin{subequations}
\begin{align}
\delta Z_{q}= & \bar{u}\left(p\right)\frac{\partial\Sigma\left(p\right)}{\partial\left(n\cdot p\right)}u\left(p\right)/\left[\bar{u}\left(p\right)n\!\!\!/u\left(p\right)\right],
\end{align}
\begin{align}
\delta Z_{\bar{q}}= & \bar{v}\left(p\right)\frac{\partial\Sigma\left(p\right)}{\partial\left(n\cdot p\right)}v\left(p\right)/\left[\bar{v}\left(p\right)n\!\!\!/v\left(p\right)\right].
\end{align}
\end{subequations}
At order $\alpha_s$, they read
\begin{subequations}
\begin{align}
\delta Z_{q}^{(1)}= & -C_{F}g_{s}^{2}\left(\frac{\mu_{\rm IR}^{2}e^{\gamma_E}}{4\pi}\right)^{\epsilon}\int
\frac{d^{4-2\epsilon}k}{\left(2\pi\right)^{4-2\epsilon}}\bar{u}
\left(p\right)\frac{1}{k\!\!\!/+p\!\!\!/-m+i\epsilon}n\!\!\!/
\nonumber \\
 & \times \frac{1}{k\!\!\!/+p\!\!\!/-m+i\epsilon}u\left(p\right)D_{\mu\nu}\left(k\right)/\left[\bar{u}\left(p\right)n\!\!\!/u\left(p\right)\right],
\end{align}
\begin{align}
\delta Z_{\bar q}^{(1)}= & - C_{F} g_{s}^{2}\left(\frac{\mu_{\rm IR}^{2}e^{\gamma_E}}{4\pi}\right)^{\epsilon}\int
\frac{d^{4-2\epsilon}k}{\left(2\pi\right)^{4-2\epsilon}}\bar{v}
\left(p\right)\frac{1}{k\!\!\!/-p\!\!\!/-m+i\epsilon}n\!\!\!/
\nonumber \\
 & \times\frac{1}{k\!\!\!/-p\!\!\!/-m+i\epsilon}v\left(p\right)D_{\mu\nu}\left(k\right)/\left[\bar{v}\left(p\right)n\!\!\!/v\left(p\right)\right].
\end{align}
\label{eq:WvFncRnml_1loop}
\end{subequations}
The detailed derivation of their analytic expressions is also presented in the Appendix~\ref{1loopExmpl}.

Both the vertex diagram and self-energy diagrams contain IR divergences.
After some manipulations as elaborated in Appendix~\ref{1loopExmpl},
we are able to isolate those IR divergent parts as the terms containing $\left(1-2x\right)^{-1-2\epsilon}$ and
$\left(1-2x\right)^{-2-2\epsilon}$. With the aid of the distribution identities listed in Appendix~\ref{Distribution:Identities},
we can rewrite these terms as the IR pole of the form $\delta(x-1/2)/\epsilon_{IR}$ and the
plus (double-plus) functions. The ``+'' and ``++'' functions are distributions, in the sense that when convoluted with a test function $g(x)$, which give
\begin{subequations}
\begin{align}
\int_{0}^{\frac{1}{2}}dx\,\left[f\left(x\right)\right]_{+}g\left(x\right) & =\int_{0}^{\frac{1}{2}}dx\,f\left(x\right)\left[g\left(x\right)-g\left(\frac{1}{2}\right)\right],\\
\int_{0}^{\frac{1}{2}}dx\,\left[f\left(x\right)\right]_{++}g\left(x\right) & =\int_{0}^{\frac{1}{2}}dx\,f\left(x\right)\left[g\left(x\right)-g'\left(\frac{1}{2}\right)
\left(x-\frac{1}{2}\right)-g\left(\frac{1}{2}\right)\right],
\end{align}
\end{subequations}
where $g\left(x\right)$ is regular at $x=1/2$. The above two definitions are also valid if we replace the integration range
from $0\leq x \leq 1/2$ to $1/2 \leq x \leq 1$.

Upon summing the vertex and self-energy diagrams, all the double and single IR poles cancel, and we are left with regular (at $x=1/2$) functions as well as plus distributions. After some reshuffling of terms, we are able to rewrite $\Phi^{(1)}_{c\bar{c}}(x)$ ($\widetilde{\Phi}^{(1)}_{c\bar{c}}(x)$) as an entire ``++'' function plus a IR-finite piece proportional to $\delta\left(x-\tfrac{1}{2}\right)$.

The partonic DA $\Phi^{(1)}(x)$ is not the desired short-distance distributions.
In compliance with (\ref{strategy:deter:one:loop:DA}),
we have to subtract the order-$\alpha_s$ correction to the decay constant in order to acquire the
normalized LCDA and quasi-DA.
It is straightforward to compute the order-$\alpha_s$ correction to the decay constants associated with various
$S$-wave quarkonia:
\begin{subequations}
\begin{align}
\mathfrak{f}_{\eta_{c}}^{(1)}=&	-1-\frac{\Lambda}{m}\tan^{-1}\left(\frac{m}{\Lambda}\right)
+\frac{\Lambda^{2}}{2m^{2}}\ln\left(\frac{\Lambda^{2}+m^{2}}{\Lambda^{2}}\right),
\\
\mathfrak{f}_{J/\psi}^{\parallel(1)}= & - 1-\left(\frac{\Lambda}{m}+\frac{m}{\Lambda}\right)
\tan^{-1}\left(\frac{m}{\Lambda}\right),\\
\mathfrak{f}_{J/\psi}^{\perp(1)}=& -1-\left(\frac{\Lambda}{m}+\frac{m}{\Lambda}\right)
\tan^{-1}\left(\frac{m}{\Lambda}\right)+\frac{\Lambda^{2}+m^{2}}{4m^{2}}\ln\left(\frac{\Lambda^{2}+m^{2}}{m^{2}}\right)
-\frac{\Lambda^{2}}{2m^{2}}\ln\left(\frac{\Lambda}{m}\right).
\end{align}
\label{one:loop:corr:decay:const}
\end{subequations}
Again we have kept a finite $\Lambda$, but take the IR regulator $\epsilon\to 0$, since they are IR finite.
For the $\eta_c$ and $J/\psi^\parallel$, the one-loop corrections are UV finite, and our results agree with the existing results in
literature once taking $\Lambda\to \infty$. For  $J/\psi^\perp$, the order-$\alpha_s$ correction is logarithmically UV divergent,
and our result agrees with the existing results that employ the DR as a UV regulator, but differs in the finite piece.

Not surprisingly, after incorporating the very one-loop corrections to the decay constants in (\ref{one:loop:corr:decay:const}),
the extra $\delta\left(x-\tfrac{1}{2}\right)$ pieces in $\Phi^{(1)}(x)$ get exactly cancelled
in (\ref{strategy:deter:one:loop:DA}).
We thereby obtain the properly normalized $\phi_H^{(1)}(x)$ ($\tilde{\phi}_H^{(1)}(x)$), in the sense that $\int^1_0 dx \phi_H^{(1)}(x)=\int^\infty_{-\infty} dx \tilde{\phi}_H^{(1)}(x)=0$.

\subsection{Analytic expressions of order-$\alpha_s$ LCDA and DA of $S$-wave quarkonia}

In this section, we present the analytical expressions for the order-$\alpha_s$ corrections to the DAs of
various helicity-states of $S$-wave quarkonia.

The LCDAs of three $S$-wave quarkonium states have rather quite compact form:
\begin{subequations}
\begin{align}
 & \phi_{\eta_{c}}^{(1)} (x;\Lambda,m)\nonumber \\
= & \left[x\left(\frac{2}{1-2x}+1\right)\log\left(\frac{\Lambda^{2}}{m^{2}\left(1-2x\right)^{2}}+1\right)+\frac{2x\Lambda^{2}}{\left(1-2x\right)^{2}\left(\Lambda^{2}+m^{2}\left(1-2x\right)^{2}\right)}+\left(x\rightarrow1-x\right)\right]_{++},
\end{align}
\begin{align}
 & \phi_{J/\psi}^{\parallel,(1)}(x;\Lambda,m)\nonumber \\
= & \left[x\left(\frac{2}{1-2x}+1\right)
\log\left(\frac{\Lambda^{2}}{m^{2}\left(1-2x\right)^{2}}+1\right)+\frac{8x^{2}\left(1-x\right)
\Lambda^{2}}{\left(1-2x\right)^{2}\left(\Lambda^{2}+m^{2}
\left(1-2x\right)^{2}\right)}+\left(x\rightarrow1-x\right)\right]_{++},
\end{align}

\begin{align}
 & \phi_{J/\psi}^{\perp,(1)}(x;\Lambda,m)\nonumber \\
= & \left[\frac{2x}{1-2x}\log\left(\frac{\Lambda^{2}}{m^{2}\left(1-2x\right)^{2}}+1\right)+
\frac{2x\Lambda^{2}}{\left(1-2x\right)^{2}\left(\Lambda^{2}+m^{2}\left(1-2x\right)^{2}\right)}
+\left(x\rightarrow1-x\right)\right]_{++},
\end{align}
\label{LCDA:res:analytic:one:loop}
\end{subequations}
all of which only have support in the range $0\leq x\leq1$.
These LCDAs are symmetric under $x\leftrightarrow1-x$,
as demanded by charge conjugation symmetry.

Note all these LCDAs contain explicit $\ln \Lambda$ dependence. This is in conformity with the celebrated Efremov-Radyushkin-Brodsky-Lepage (ERBL) evolution equation~\cite{Lepage:1979zb,Efremov:1979qk}:
\begin{align}
 {d\over d\ln\Lambda^2} \Phi_H(x;\Lambda) &= {\alpha_s C_F\over \pi}\int^1_0 dy V_{0} (x, y) \Phi_H(y,\Lambda)+{\mathcal O}(\alpha_s^2),
 \label{ERBL:evolution:eq}
\end{align}
where the evolution kernel $V_0(x,y)$ varies with different hadron helicity.
Substituting the $\phi_H^{(1)}$ in (\ref{LCDA:res:analytic:one:loop}) back to (\ref{LCDA:pert:expansion}), and plugging into  (\ref{ERBL:evolution:eq}), also taking into account the order-$\alpha_s$ correction to decay constant in (\ref{one:loop:corr:decay:const}),
it is straightforward to check that all of these LCDAs indeed obey the ERBL equation.

Conceivably, the non-logarithm terms in (\ref{LCDA:res:analytic:one:loop}) differ from those given in \cite{Wang:2013ywc}~\footnote{When computing the order-$\alpha_s$ correction to color-singlet  channel of double parton fragmentation function, the corresponding results in Ref.~\cite{Ma:2013yla} are equivalent to those of the LCDA~\cite{Wang:2013ywc}.},
which can be attributed to the different choice of the UV regulators.
Had we first sent $\Lambda\to\infty$ during the intermediate stage,
which amounts to use DR to regulate both UV and IR divergences, we would be able to reproduce their results.

Next we turn to the quasi-DA. Since the boost invariance is sacrificed there,
the expressions would explicitly depend on $P^z$, consequently become considerably more complicated:
\begin{subequations}
\begin{align}
 & \tilde{\phi}_{\eta_{c}}^{(1)}(x, P^z;\Lambda,m)\nonumber \\
= & \left[\ensuremath{-x\left(1+\frac{2}{1-2x}\right)\frac{p^{z}}{p^{0}}\ln\left(\frac{p^{0}\sqrt{\Lambda^{2}+m^{2}+4\left(p^{z}\right)^{2}x^{2}}+2\left(p^{z}\right)^{2}x+m^{2}}{p^{0}\sqrt{m^{2}+4\left(p^{z}\right)^{2}x^{2}}+2\left(p^{z}\right)^{2}x+m^{2}}\right)}\right.\nonumber \\
 & \ensuremath{+\ensuremath{\frac{(1-x)p^{z}}{p^{0}}\ln\left(\frac{(1-2x)\left(p^{z}\right)^{2}+\sqrt{\Lambda^{2}+(-1+2x)^{2}\left(p^{z}\right)^{2}}p^{0}}{\left(1-2x\right)\left(p^{z}\right)^{2}+p^{z}\sqrt{\left(p^{z}\right)^{2}+m^{2}}\left|1-2x\right|}\right)}}\nonumber \\
 & \ensuremath{-\frac{2(1-x)p^{z}}{(1-2x)p^{0}}\ln\left(\frac{\sqrt{\left(\Lambda^{2}+(1-2x)^{2}\left(p^{z}\right)^{2}\right)}p^{0}+\left(1-2x\right)\left(p^{z}\right)^{2}}{\left|1-2x\right|p^{z}p^{0}+\left(1-2x\right)\left(p^{z}\right)^{2}}\right)}\nonumber \\
 & \ensuremath{+\frac{\left(\Lambda^{2}\left(p^{0}\right)^{2}-2m^{2}x(1-2x)\left(p^{z}\right)^{2}\right)\sqrt{\Lambda^{2}+4x^{2}\left(p^{z}\right)^{2}+m^{2}}}{2\left(m^{2}(1-2x)^{2}\left(p^{z}\right)^{2}+\Lambda^{2}\left(p^{0}\right)^{2}\right)\left(1-2x\right)p^{z}}}+\ensuremath{\frac{x\sqrt{4x^{2}\left(p^{z}\right)^{2}+m^{2}}}{(1-2x)^{2}p^{z}}}\nonumber \\
 & \left.+\ensuremath{\frac{m^{2}p^{z}\sqrt{\Lambda^{2}+\left(1-2x\right)^{2}\left(p^{z}\right)^{2}}}{2\left(\Lambda^{2}+(1-2x)^{2}\left(p^{z}\right)^{2}\right)m^{2}+2\Lambda^{2}\left(p^{z}\right)^{2}}}+\frac{\sqrt{\Lambda^{2}+\left(1-2x\right)^{2}\left(p^{z}\right)^{2}}}{2\left(1-2x\right)^{2}p^{z}}-\frac{1}{\left|1-2x\right|}+\left(x\rightarrow1-x\right)\right]_{++},
\end{align}

\begin{align}
 & \tilde{\phi}_{J/\psi}^{\parallel,(1)}(x, P^z;\Lambda,m)\nonumber \\
= & \left[\!\left(\!\frac{p^{z}\!\left(m^{2}\!\!+\!2x\left(p^{z}\right)^{2}\right)}
{2\left(p^{0}\right)^{3}}\!+\!\frac{p^{z}\!\left(m^{2}(2x\!+\!1\!)\!+\!4x\left(p^{z}\right)^{2}\right)}{2(1\!-\!2x\!)\!\left(p^{0}\right)^{3}}\!\right)\!\ln\!\left(\!\!\frac{p^{0}\sqrt{m^{2}\!+\!\Lambda^{2}\!+\!4x^{2}\!\left(p^{z}\!\right)^{2}}\!-\!m^{2}\!-\!2x\!\left(p^{z}\right)^{2}}{p^{0}\sqrt{m^{2}\!+\!4x^{2}\left(p^{z}\right)^{2}}\!-\!m^{2}\!-\!2x\!\left(p^{z}\right)^{2}}\!\right)\right.\nonumber \\
 & +\!\left(\!\frac{(1\!-\!2x)\left(p^{z}\right)^{3}}{2\left(p^{0}\right)^{3}}\!-\!\frac{p^{z}\!\left(m^{2}(2x\!+\!1\!)\!+\!4x\left(p^{z}\right)^{2}\right)}{2(1\!-\!2x)\!\left(p^{0}\right)^{3}}\!\right)\!\ln\!\left(\!\!\frac{p^{0}\sqrt{\Lambda^{2}\!+\!(1\!-\!2x)^{2}\left(p^{z}\!\right)^{2}}\!+\!(1\!-\!2x\!)\left(p^{z}\right)^{2}\!}{\!p^{z}p^{0}\left|1-2x\right|+(1\!-\!2x\!)\left(p^{z}\right)^{2}}\right)\nonumber \\
 & -\!\frac{p^{z}}{4p^{0}}\!\log\!\left(\!\frac{\Lambda^{2}\left(p^{0}\right)^{2}}{m^{2}\left(p^{z}\right)^{2}(1-2x)^{2}}\!+\!1\!\right)\!+\!\frac{\left(m^{2}\!+\!4(1\!-\!x)x\left(p^{z}\right)^{2}\right)\left(\!\sqrt{m^{2}\!+\!4x^{2}\left(p^{z}\right)^{2}}\!-\!|1-2x|p^{z}\!\right)}{2(1\!-\!2x)^{2}p^{z}\left(p^{0}\right)^{2}}\nonumber \\
 & \ensuremath{+\!\frac{m^{2}p^{z}\left(m^{2}-4(x-1)x\left(p^{z}\right)^{2}\right)\left(\sqrt{\Lambda^{2}\!+\!(1-2x)^{2}\left(p^{z}\right)^{2}}\!-\!\sqrt{m^{2}\!+\!\Lambda^{2}\!+\!4x^{2}\left(p^{z}\right)^{2}}\right)}{2\left(p^{0}\right)^{2}\left(m^{2}\left(\Lambda^{2}+(1-2x)^{2}\left(p^{z}\right)^{2}\right)+\Lambda^{2}\left(p^{z}\right)^{2}\right)}}\nonumber \\
 & \left.+\!\frac{p^{z}\left(\sqrt{m^{2}\!+\!\Lambda^{2}\!+\!4x^{2}\left(p^{z}\right)^{2}}\!-\!\sqrt{m^{2}\!+\!4x^{2}\!\left(p^{z}\!\right)^{2}}\right)}{2(1-2x)\left(p^{0}\right)^{2}}\!+\!\frac{\sqrt{\Lambda^{2}\!+\!\left(\!1\!-\!2x\!\right)^{2}\left(p^{z}\!\right)^{2}}}{2\left(1-2x\right)^{2}p^{z}}\!-\!\frac{1}{2\left|1-2x\right|}\!+\!\left(x\rightarrow\!1\!-\!x\!\right)\!\right]_{++},
\end{align}

\begin{align}
 & \tilde{\phi}_{J/\psi}^{\perp,(1)}(x, P^z;\Lambda,m)\nonumber \\
= & \left[-\frac{p^{z}}{p^{0}}\!\ln\!\left(\frac{p^{0}\sqrt{\Lambda^{2}+(1\!-\!2x)^{2}\left(p^{z}\right)^{2}}+(1\!-\!2x\!)\left(p^{z}\right)^{2}\!}{\!p^{z}p^{0}\left|1-2x\right|+(1\!-\!2x\!)\left(p^{z}\right)^{2}}\right)\right.\nonumber \\
 & -\frac{2xp^{z}}{p^{0}}\ln\left(\frac{p^{0}\sqrt{\Lambda^{2}+m^{2}+4\left(p^{z}\right)^{2}x^{2}}+2\left(p^{z}\right)^{2}x+m^{2}}{p^{0}\sqrt{m^{2}+4\left(p^{z}\right)^{2}x^{2}}+2\left(p^{z}\right)^{2}x+m^{2}}\right)\nonumber \\
 & -\!\frac{p^{z}}{2p^{0}}\!\ln\!\left(\!\frac{\Lambda^{2}\left(p^{0}\right)^{2}}{m^{2}(1-2x)^{2}\left(p^{z}\right)^{2}}\!+\!1\!\right)+\!\frac{\sqrt{\Lambda^{2}\!+m^{2}\!+\!4x^{2}\left(p^{z}\right)^{2}}}{2\left(1-2x\right)p^{z}}\nonumber \\
 & \ensuremath{-\frac{m^{2}p^{z}\sqrt{\Lambda^{2}\!+m^{2}\!+\!4x^{2}\left(p^{z}\right)^{2}}}{2m^{2}\left(p^{z}\right)^{2}\left(1-2x\right)^{2}+2\left(p^{0}\right)^{2}\Lambda^{2}}}+\frac{x\sqrt{m^{2}\!+\!4x^{2}\left(p^{z}\right)^{2}}}{p^{z}\left(1-2x\right)^{2}}-\frac{1}{\left|1-2x\right|}\nonumber \\
 & \left.+\frac{m^{2}p^{z}\sqrt{\Lambda^{2}\!+\!(1-2x)^{2}\left(p^{z}\right)^{2}}}{2\left(m^{2}
 \left(p^{z}\right)^{2}\left(1-2x\right)^{2}+\left(p^{0}\right)^{2}\Lambda^{2}\right)}+
 \frac{\sqrt{\Lambda^{2}\!+\!(1-2x)^{2}\left(p^{z}\right)^{2}}}{2\left(1-2x\right)^{2}p^{z}}+
 \left(x\rightarrow1-x\right)\right]_{++},
\end{align}
\label{quasi-DA:res:analytic:one:loop}
\end{subequations}
where $p^{0}\equiv \sqrt{\left(p_z\right)^{2}+m^{2}}$.
Needless to say, these quasi-DAs are also symmetric under the transformation
$x\leftrightarrow 1-x$.

A major difference between quasi DAs  and LCDA is that the former has a nonvanishing support when
$x<0$ and $x>1$, though suppressed by $1/\left(P^{z}\right)^{2}$.
This is a general feature of quasi-distributions, which was also seen in the
quasi PDF and GDP. This simply signals the breakdown of a physical parton interpretation
for quasi distributions.

Reassuringly, it can be analytically checked, when boosted to the IMF by sending $P^{z}\rightarrow\infty$,
these frame-dependent quasi-DAs in (\ref{quasi-DA:res:analytic:one:loop}) indeed reduce to
the exact functional form of the LCDAs in (\ref{LCDA:res:analytic:one:loop}).

Note all the $\tilde{\phi}_H^{(1)}(x)$ in (\ref{LCDA:res:analytic:one:loop}) contain a linearly UV-divergent piece $\sqrt{\Lambda^{2}+\left(1-2x\right)^{2}\left(p^{z}\right)^{2}}$.
This term is always accompanied with the double pole $\left(1-2x\right)^{-2}$, and suppressed by $1/P^z$.
Physically, such term can be traced to the self-energy correction to the gauge link.
In this work, we have keep $\Lambda$ finite, so such a term does not bring any problem.
For a consistent renormalization program, the proper treatment of such term in the limit $\Lambda\to \infty$ at fixed $P^z$
has been sketched in Ref.~\cite{Xiong:2013bka,Ji:2015qla}.

Had we taken the $\Lambda \gg P^z\gg m$ limit in quasi-DA, the BL evolution kernel would also emerge
in the range $0\leq x\leq 1$, but the accompanied logarithm is in the form of $\ln P^z$ rather than $\ln \Lambda$ as in LCDA.
It is the strength of the recently advocated LaMET~\cite{Ji:2014gla} to resum such type
of logarithms.

\section{Numerical comparison between LCDA and Quasi-DA of quarkonia}
\label{Numerical:study:DA}

Having the one-loop exact ``data'' available for both LCDA and quasi-DA of the $\eta_c$ and $J/\psi$,
it is the time to make a comprehensive study on their properties.

\subsection{The convergence behavior of quasi-DA to LCDA with increasing $P^z$}

\begin{figure}[tbH]
\centering{}\includegraphics[scale=0.5]{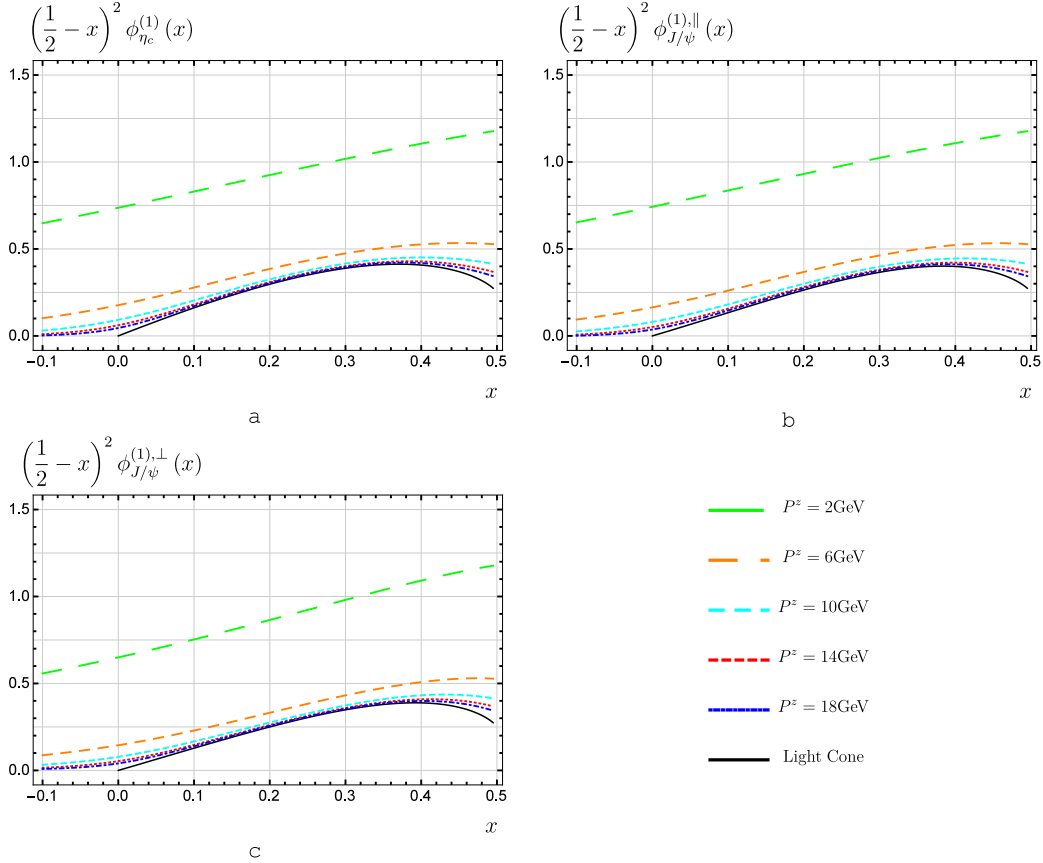}
\protect\caption{LCDAs and quasi-DAs for three $S$-wave charmonium helicity states with various $P^z$.}
\label{DA_Num}
\end{figure}

We have already seen that the quasi-DAs in the limit $P^{z}\rightarrow\infty$  analytically recover the LCDAs for each species of quarkonia. It is of practical curiosity to see explicitly how fast the quasi-DA approaches the LCDA
with increasing $P^{z}$.

For numerical study, we take the charm quark mass as $1.4$ GeV.
We tentatively choose $\Lambda=3$ GeV, approximately equal to
the masses of  $\eta_{c}$ and $J/\psi$.
In Fig.~\ref{DA_Num}, for each species of $S$-wave charmonia,
we display several sets of the quasi-DAs at various values of $P^z$:
2, 6, 10, 14, 18 GeV, respectively.
Because the DAs are symmetric under $x\leftrightarrow 1-x$, and quasi-DAs
decrease very rapidly in the unphysical regions $x<0$ and $x>1$,
we only plot them in the interval $-0.1<x<1/2$.
In order to suppress the singular appearance near $x \sim{1\over 2}$,
we deliberately multiply all the DAs by $(1-2x)^{2}$.

From Fig.~\ref{DA_Num}, we clearly see the trend of quasi-DA approaching the LCDA with increasing $P^z$.
Also, it is interesting to note that the quasi-DA admits a nonzero value at $x=0$ and 1, which
is quite different from LCDA.
\begin{figure}[tbH]
\centering{}\includegraphics[scale=0.5]{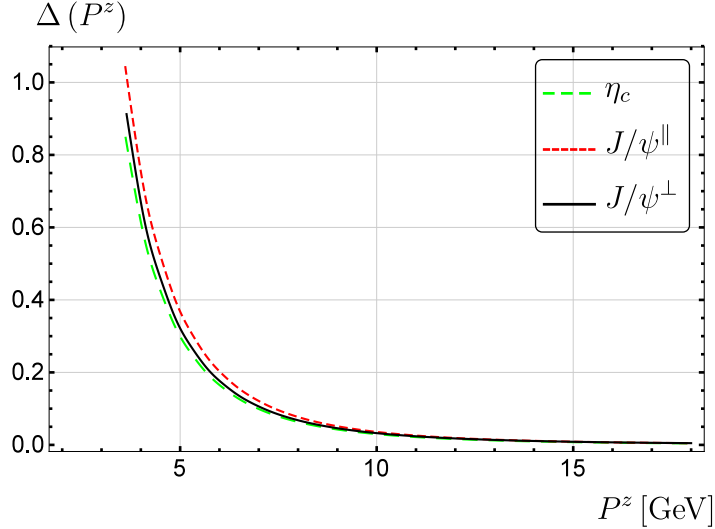}\protect\caption{The degree of resemblance
as a function of $P^{z}$.}
\label{Rel_Distance}
\end{figure}

In order to quantify the difference between LCDA and quasi-DA at a given $P^z$,
we invent a parameter {\it degree of resemblance}, denoted by $\Delta(P^z)$:
\begin{align}
\Delta_H(P^{z})= & \frac{\int_{0}^{\frac{1}{2}}dx\,\left(1-2x\right)^{4}\left[\phi_H^{(1)}\left(x;\Lambda,m\right)-
\tilde{\phi}_H^{(1)}(x,P^z;\Lambda,m)\right]^{2}}{\int_{0}^{\frac{1}{2}}dx\,\left(1-2x\right)^{4}
\left[\phi_H^{(1)}\left(x;\Lambda,m\right)\right]^2}.
\end{align}
The dependence of $\Delta_H$  on $P^{z}$ is shown in Fig.~\ref{Rel_Distance}.
It is a rapidly descending function.
When $P^{z}$ is 6 GeV, $\Delta$ is about $20\%$; as $P^{z}$ is boosted to 9 GeV,
$\Delta$ already decreases to $5\%$.
It may persuasively imply that, provided that $P^z$ is about three times larger than the hadron mass
(with renormalization scale fixed around the hadron mass), the quasi-DA already converges to the
``true'' LCDA to a decent extent. Interestingly, a very recent investigation on the
nucleon quasi-PDF in the diquark model has drawn a similar conclusion~\cite{Gamberg:2014zwa}.

\begin{figure}[tbH]
\centering{}\includegraphics[scale=0.5]{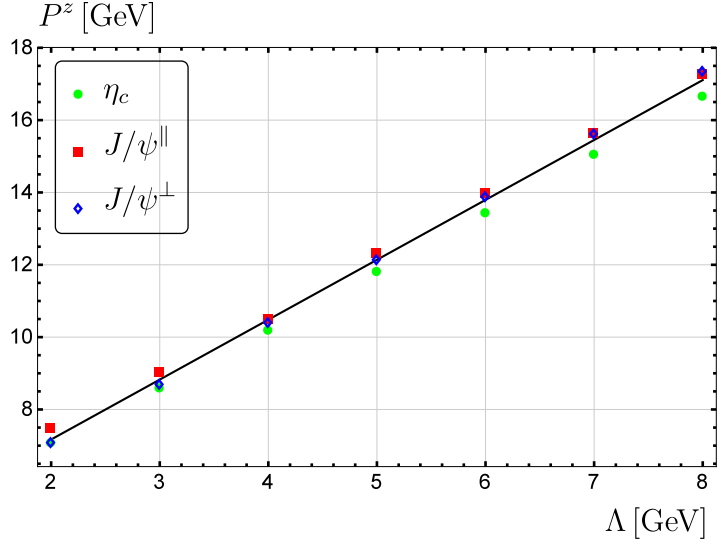}\protect\caption{
To fulfill $\Delta=0.05$, the minimal value of $P^{z}$ required as a function of $\Lambda$.}
\label{PzVsCutOff}
\end{figure}

We can further inspect the correlation between $P^{z}$ and $\Lambda$, for a given
degree of resemblance. In Fig.~\ref{PzVsCutOff}, we show that to achieve $\Delta=0.05$,
how the minimal value of $P^z$ required
depends on the value of $\Lambda$.
The colored dots/squares are generated from actual calculation,
and the solid line is a linear fit by averaging over three types of charmonium helicity states, and we obtain
$\overline{P^z}=1.66\,\Lambda+3.86$ GeV.
It is interesting to observe this linear correlation between $P^z$ and $\Lambda$.

\subsection{Comparison of first inverse moment between LCDA and quasi-DA}

In the hard exclusive reactions, the factorization theorem expresses the amplitude as
the convolution of the hard-scattering kernel with the LCDAs.
For a leading-twist contribution, the hard part always bears the form ${1\over x}$,
thereby, it is the first inverse moment of the LCDA that is of ubiquitous
phenomenological interest~\cite{Lepage:1980fj}.

We are curious to which extent the inverse moment generated from quasi-DA will resemble
the ``true'' one in magnitude. The first inverse moment of the charmonia LCDA, to the order-$\alpha_s$ accuracy,
is given by
\begin{align}
\left\langle {x^{-1}}\right\rangle_{H} &  \equiv \int_{0}^{1} \!dx \, {\phi_H^{(1)}(x;\Lambda,m) \over x}.
\end{align}
As dictated by the general principle, $\phi_H \propto x$ as $x\to 0$,
thereby the leading-twist LCDA admits a finite inverse moment.
One can readily deduce the closed form for the inverse moments from (\ref{LCDA:res:analytic:one:loop}),
\begin{subequations}
\begin{align}
\left\langle x^{-1}\right\rangle _{\eta_{c}}= & -\frac{\pi^{2}}{6}+\frac{2\Lambda^{2}\ln2}{m^{2}+\Lambda^{2}}+\frac{2\Lambda\left(2m^{2}+\Lambda^{2}\right)}
{m\left(m^{2}+\Lambda^{2}\right)}\arctan\frac{m}{\Lambda}-\ln 2 \ln{\Lambda^2+m^2\over \Lambda^2}
\nonumber \\
 & +{\Lambda^4+2\Lambda^2 m^2+3m^4\over 2m^2(\Lambda^2+m^2)} \ln{\Lambda^2+m^2\over \Lambda^2}+ (3-2\ln 2)\ln{\Lambda\over m}
 \nonumber \\
 & + 2\,{\rm Re} \left[{\rm Li}_{2}\left(\frac{2m}{m-i\Lambda}\right)-{\rm Li}_{2}\left(\frac{m}{m-i\Lambda}\right)\right],
\end{align}
\begin{align}
\left\langle x^{-1}\right\rangle _{J/\psi^{\parallel}}= & -\frac{\pi^{2}}{6}+\frac{4\Lambda}{m}\arctan\frac{m}{\Lambda}
-{\Lambda^2-(3-2\ln2)m^2 \over 2m^2} \ln{\Lambda^2+m^2\over \Lambda^2}
 \nonumber \\
& +(3-2\ln 2)\ln{\Lambda\over m}+ 2\,{\rm Re} \left[{\rm Li}_{2}\left(\frac{2m}{m-i\Lambda}\right)-{\rm Li}_{2}\left(\frac{m}{m-i\Lambda}\right)\right],
\end{align}
\begin{align}
\left\langle x^{-1}\right\rangle _{J/\psi^{\perp}}= & -\frac{\pi^{2}}{3}+\frac{2\Lambda^{2}\ln2}{m^{2}+\Lambda^{2}}+\frac{2\Lambda\left(3m^{2}+2\Lambda^{2}\right)}
{m\left(m^{2}+\Lambda^{2}\right)}\arctan\frac{m}{\Lambda}
\nonumber \\
 & + 4(1-\ln 2)\ln{\Lambda\over m} -{(2\ln 2-1)\Lambda^2+2(\ln2-1)m^2\over \Lambda^2+m^2} \ln{\Lambda^2+m^2\over \Lambda^2}
 \nonumber \\
 & + 4\,{\rm Re} \left[{\rm Li}_{2}\left(\frac{2m}{m-i\Lambda}\right)-{\rm Li}_{2}\left(\frac{m}{m-i\Lambda}\right)\right].
\end{align}
\end{subequations}
In the $\Lambda\gg m$ limit, the inverse moments are dominated by the $\ln {\Lambda\over m}$ term for each LCDA,
whose coefficients agree with the previously known results~\cite{Jia:2010fw,Wang:2013ywc}.
It is this type of collinear logarithms that can be resummed to orders in
$\alpha_s$ with the aid of ERBL evolution equation~\cite{Jia:2008ep}.

In contrast, as can be analytically inferred from (\ref{quasi-DA:res:analytic:one:loop}),
or clearly seen from Fig.~\ref{DA_Num}, the quasi-DAs approach nonzero values as $x\to 0$.
Nevertheless, the quasi-DAs still smoothly cross $x=0$,
we thus utilize the principal value prescription to
define the inverse moment of the quasi-DA:
\begin{align}
\left\langle {\tilde x^{-1}}\right\rangle_{H} &  \equiv {\rm P.~V.}\,\int_{-\infty}^{\infty} \!dx \, {\tilde{\phi}_H^{(1)}(x,P^z;\Lambda,m) \over x}.
\end{align}

\begin{figure}[H]
\centering{}\includegraphics[scale=0.5]{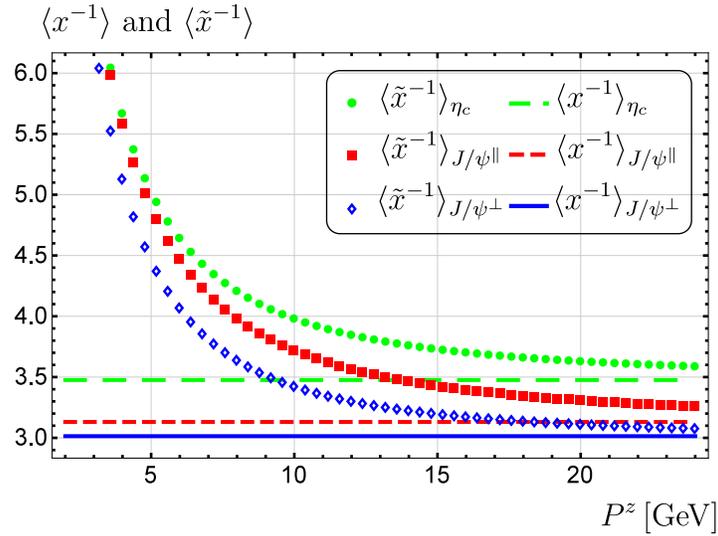}\protect\caption{The first inverse moments obtained from various $S$-wave quarkonia, for both order-$\alpha_s$ LCDA and quasi-DA, as function of $P^z$. We fix $\Lambda=3$ GeV.}
\label{frst_inv_mnts}
\end{figure}

The magnitudes of the first inverse moments of LCDA and quasi-DA as a function of $P^z$
are presented in Fig.~\ref{frst_inv_mnts}, with $\Lambda$ fixed at 3 GeV.
To characterize the extent of the proximity of the first
inverse moments between LCDA and quasi-DA, we introduce the following fractional difference:
\begin{equation}
\Omega_H\left(P^{z},\Lambda\right)=\left|\frac{\langle x^{-1}\rangle_H-
\langle\tilde{x}^{-1}\rangle_H}
{\langle x^{-1}\rangle_H} \right|.
\end{equation}
Concretely speaking, at $\Lambda=3$ GeV, the $\Omega_H$ for three types of $S$-wave states are
\begin{center}
\begin{tabular}{|c|c|c|c|}
\hline
\;$P^z$\; & \;$\Omega_{\eta_{c}}$\; & \;$\Omega_{J/\psi^{\parallel}}$\; & \;$\Omega_{J/\psi^{\perp}}$\;\tabularnewline
\hline
\hline
\;6  GeV\; & \;$0.335$\; & \;$0.431$\; & \;$0.358$\;\tabularnewline
\hline
\;9 GeV\; & \;$0.172$\; & \;$0.228$\; & \;$0.174$\;\tabularnewline
\hline
\;18 GeV\; & \;$0.052$\; & \;$0.071$\; & \;$0.049$\;\tabularnewline
\hline
\end{tabular}.
\par\end{center}
We see that, even when $P^z$ is boosted to 9 GeV, the inverse moments generated from the quasi-DA still differ
from the true results about 20\%.
Compared with $\Delta_H(P^{z})$,  the first inverse moments generated by the quasi-DAs appear
to approach the LCDA value with a rather slower pace.

\section{Summary}
\label{summary}

PDF and LCDA are among the most prominent and basic nonperturbative quantities
coined in QCD, encapsulating rich dynamics about the quark-gluon degree of freedom inside a hadron.
For several decades, how to effectively compute such light-cone distributions from the first principle of QCD
has posed a preeminent challenge and the progress was slow.
As a breakthrough, the recently advocated quasi distributions, and the corresponding LaMET,
have the very bright prospect to help finally overcome this long-standing difficulty.

Despite some important progress, there remain eminent technical obstacles for lattice to make
phenomenologically competitive measurements on the quasi distributions.
One is the lack of systematic renormalization program for the quasi distributions.
Moreover, for the current lattice technique, to make a precise simulation for the quasi distributions
in a highly boosted hadron state is also unrealistic.
Therefore, it is valuable if some useful
insights about the general aspects of the quasi distributions can be gained in the
continuum field theory.

Thank to its tremendous simplicity, heavy quarkonium actually provides an ideal theoretical laboratory
to study the quasi distributions. It has been known that NRQCD factorization allows one to express
the LCDA of a heavy quarkonium,
at the lowest order in velocity expansion, simply as the product of a perturbatively-calculable, IR-finite coefficient function
and a single nonperturbative matrix element.
Therefore, the profile of the quarkonium LCDA is fully amenable to perturbation theory.
Quarkonium thus constitutes a rare example that the light-cone correlators
can be fairly well understood in the continuum theory without much effort.

In this work, extending the previous work on quarkonium LCDA, we apply NRQCD factorization further to
the quasi-DA of the ground-state quarkonia, and calcualte the respective coefficient functions to order $\alpha_s$.
We are able to show analytically that, the quasi-DA exactly reduces into LCDA in the infinity-momentum limit.
We also observe that, provided that the $P^z$ of a charmonium  is about 2-3 times its mass, and
with the renormalization scale chosen around the charmonium mass,
the respective quasi-DAs will converge to the LCDAs to a satisfactory degree.

Our work also has some limitation, chiefly in that we have resided entirely in a cutoff theory
and naively interpreted the hard transverse momentum cutoff $\Lambda$ as the
renormalization scale. It is worth pursuing the rigorous renormalization procedure to the quasi DAs
in the future work.

We hope our comprehensive study of the quasi-DAs for heavy quarkonia
will provide some useful guidance to the future lattice investigation
of similar quasi distributions, {\it e.g.}, how to optimally choose the parameters in their Monte Carlo simulation.

\vspace{0.3 cm}
{\noindent \it Acknowledgment.}
We thank Deshan Yang for useful discussions.
The work of Y.~J. is supported in part by the National Natural Science Foundation of China under Grants No.~11475188,
No.~11261130311 (CRC110 by DGF and NSFC), by the IHEP Innovation Grant under contract number Y4545170Y2,
and by the State Key Lab for Electronics and Particle Detectors.

\appendix

\section{Details of conducting the one-loop calculation}
\label{1loopExmpl}

It is easiest to compute the order-$\alpha_s$ correction to DAs of
the longitudinally-polarized $J/\psi$.
Therefore, in this section, we take the Feynman part (the $g_{\mu\nu}$ part in the gluon propagator in (\ref{def:gluon:propagator:axial:gauge}))
as an concrete example, to illustrate the intermediate steps in the one-loop calculation for the
LCDA and quasi-DA of $J/\psi^\parallel$.

\subsection{LCDA} \label{Appendix:details:lcda}

For the LCDA of $J/\psi^{\parallel}$, we take
$n^{\mu}$ as the light-like reference vector defined in
(\ref{reference:vector:n:def}).
The Feynman part of the vertex diagram in (\ref{eq:DA:one:loop:vertex}) reads
\begin{align}
\mathcal{I}\left(x\right)= & \frac{C_{F}}{\left(2\pi\right)^{4-2\epsilon}}\left(\frac{\mu_{\rm IR}^{2}e^{\gamma_E}}{4\pi}\right)^{\epsilon}\int d^{d}k\,\bar{u}\left(p\right)\left(-ig_{s}\gamma^{\mu}\right)\frac{i}{p\!\!\!/+k\!\!\!/-m+i\epsilon}
\gamma^{+}\frac{i}{p\!\!\!/-k\!\!\!/-m+i\epsilon}\nonumber \\
 & \times\left(-ig_{s}\gamma^{\nu}\right)v\left(p\right)\frac{-ig_{\mu\nu}}{k^{2}+i\epsilon}\delta\left(x-\frac{1}{2}-\frac{k^{+}}{2p^{+}}\right)/\bar{u}\left(p\right)\gamma^{+}v\left(p\right)\nonumber \\
 & \frac{-ig_{s}^{2}C_{F}}{\left(2\pi\right)^{4-2\epsilon}}\left(\frac{\mu_{\rm IR}^{2}e^{\gamma_E}}{4\pi}\right)^{\epsilon}\frac{2\pi^{1-\epsilon}}{\Gamma\left(1-\epsilon\right)}\int_{0}^{\Lambda}dk_{\perp}\,k_{\perp}^{1-2\epsilon}\int dk^{-}dk^{+}\nonumber \\
 & \times\frac{-4\left(1-\epsilon\right)k_{\perp}^{2}+\frac{8p^{-}}{p^{+}}\left(k^{+}\right)^{2}-8m^{2}}{\left[2\left(k^{-}+p^{-}\right)\left(k^{+}+p^{+}\right)-k_{\perp}^{2}-m^{2}+i\epsilon\right]\left[2\left(k^{-}-p^{-}\right)\left(k^{+}-p^{+}\right)-k_{\perp}^{2}-m^{2}+i\epsilon\right]}\nonumber \\
 & \times\frac{\delta\left(x-\frac{1}{2}-\frac{k^{+}}{2p^{+}}\right)
 \bar{u}\left(p\right)\gamma^{+}v\left(p\right)}{2k^{-}k^{+}-k_{\perp}^{2}+i\epsilon}.
\end{align}
The $k^{-}$ integral is carried out by contour integration.
The $\delta$-function trades the $k^{+}$ in favor of the dimensionless
momentum fraction $x$. As is well known, the $\mathcal{I}(x)$ vanishes unless when $0<x<1$,
for which the poles are distributed in both upper and lower half of the complex plane when carrying out the $k^{-}$ integration.
After utilizing Cauchy's theorem, we are left with the integration over the transversa momentum:
\begin{align}
\mathcal{I}\left(x\right)= & \frac{g_{s}^{2} C_{F}}{\left(2\pi\right)^{3-2\epsilon}}\left(\frac{\mu_{\rm IR}^{2}e^{\gamma_E}}{4\pi}\right)^{\epsilon}\frac{4\pi^{1-\epsilon}}{\Gamma\left(1-\epsilon\right)}
\int_{0}^{\Lambda}dk_{\perp}\,k_{\perp}^{1-2\epsilon}\theta\left(x\right)\theta\left(\frac{1}{2}-x\right)x\nonumber \\
 & \times\frac{\left(1-\epsilon\right)k_{\perp}^{2}+m^{2}\left(4\left(1-\epsilon\right)x\left(1-x\right)
 +1+\epsilon\right)}{\left[k_{\perp}^{2}+m^{2}\left(1-2x\right)^{2}\right]^{2}}\nonumber \\
= & \left\{ \frac{2\theta\left(x\right)\theta\left(\frac{1}{2}-x\right)e^{\epsilon\gamma}x
\left(1-\epsilon\right){}^{4}\Lambda^{4-2\epsilon}\mu_{\rm IR}^{2\epsilon}}{\left(2-\epsilon\right)\Gamma\left(1-\epsilon\right)\left(1-2x\right)^{4}m^{4}}
\left[\left(1-\epsilon\right){}_{2}F_{1}\left(1,2-\epsilon;3-\epsilon;-\frac{\Lambda^{2}}{m^{2}
\left(1-2x\right)^{2}}\right)\right.\right.\nonumber \\
 & \left.\left.+\frac{\left(\epsilon-2\right)m^{2}\left(1-2x\right)^{2}}{m^{2}\left(1-2x\right)^{2}+\Lambda^{2}}\right]\right\} _{\text{1st}}\nonumber \\
 & +\left\{ \frac{2\theta\left(x\right)\theta\left(\frac{1}{2}-x\right)e^{\epsilon\gamma_{E}}x
 \left(4\left(1-\epsilon\right)x\left(1-x\right)+1+\epsilon\right)
 {}^{4}\Lambda^{-2\epsilon}\mu_{\rm IR}^{2\epsilon}}{\left(1-2x\right)^{4}}\right.\nonumber \\
 & \left.\times\left[\frac{\Lambda^{2}\left(1-2x\right)^{2}}{\left(m^{2}\left(1\!-\!2x\right)^{2}\!+
 \!\Lambda^{2}\right)\Gamma\left(1\!-\!\epsilon\right)}\!+\!\frac{\epsilon\Lambda^{2}}{m^{2}\Gamma\left(2\!-
 \!\epsilon\right)}\,{}_{2}F_{1}\left(\!1,1\!-\!\epsilon;2\!-\!\epsilon;\!-\!\frac{\Lambda^{2}}
 {m^{2}\left(1\!-\!2x\right)^{2}}\!\right)\right]\right\} _{\text{2nd}}\nonumber \\
+ & \left(x\rightarrow1-x\right).
\label{I:vertex:feyn:part:jpsi:long}
\end{align}

For the first piece $\mathcal{I}_{1}(x)$ (with the subscript ``1st''),
which originates from the part of the integrand containing $k_{\perp}^{2}$,
one can safely set $\epsilon\to 0$, because it is regular at
$x={1\over 2}$. The result is
\begin{align}
\mathcal{I}_{1}\left(x\right)= & \frac{C_{F}g_{s}^{2}}{4\pi^{2}}\theta\left(x\right)\theta\left(\frac{1}{2}-x\right)x
\left[\ln\left(1+\frac{\Lambda^{2}}{m^{2}\left(1-2x\right)^{2}}\right)-\frac{\Lambda^{2}}
{\Lambda^{2}+m^{2}\left(1-2x\right)^{2}}\right].\nonumber \\
 & +\left(x\rightarrow1-x\right).
\end{align}

The second piece $\mathcal{I}_{2}(x)$ in (\ref{I:vertex:feyn:part:jpsi:long}) (labelled by the subscript ``2nd'')
turns out to be IR divergent at $x={1\over 2}$.
To isolate the IR pole, we first expand the hypergeometric function
to $\mathcal{O}(\epsilon^{1})$:
\begin{align}
\mathcal{I}_{2}\left(x\right)= & \frac{C_{F}g_{s}^{2}}{4\pi^{2}}\frac{\theta\left(x\right)\theta\left(\frac{1}{2}-x\right) e^{\epsilon\gamma_E} x\left(4\left(1-\epsilon\right)
x\left(1-x\right)+1+\epsilon\right)\Lambda^{-2\epsilon}\mu_{\rm IR}^{2\epsilon}}{\left(1-2x\right)^{2}}\nonumber \\
 & \times\left[\frac{\Lambda^{2}}{\left(\Lambda^{2}\!+\!m^{2}\left(1\!-\!2x\right)^{2}\right)\Gamma\left(1\!-\!\epsilon\right)}\!+\!\epsilon\ln\left(\!\frac{\Lambda^{2}\!+\!m^{2}\left(1\!-\!2x\right)^{2}}{m^{2}}\!\right)-2\epsilon\ln\left(1-2x\right)\right]\nonumber \\
 & +\mathcal{O}\left(\epsilon^{1}\right).
\end{align}

Using the relation
\begin{align}
\left(\frac{1}{1-2x}\right)^{2+2\epsilon}-\frac{1-2\epsilon\ln\left(1-2x\right)}{\left(1-2x\right)^{2}} & =\mathcal{O}\left(\epsilon^{2}\right),
\end{align}
we can rewrite $\mathcal{I}_{2}(x)$ as
\begin{align}
\mathcal{I}_{2}\left(x\right)= & \frac{C_{F}g_{s}^{2}}{4\pi^{2}}\frac{\theta\left(x\right)\theta\left(\frac{1}{2}-x\right)e^{\epsilon\gamma}x\left(4\left(1-\epsilon\right)x\left(1-x\right)+1+\epsilon\right)\Lambda^{-2\epsilon}
\mu_{\rm IR}^{2\epsilon}}{\Gamma\left(1-\epsilon\right)}\nonumber \\
 & \times\left[-\frac{m^{2}}{m^{2}\left(1-2x\right)^{2}+\Lambda^{2}}+
 \left(\frac{1}{1-2x}\right)^{2+2\epsilon}\left(1+\epsilon\ln\left(\frac{\Lambda^{2}+
 m^{2}\left(1-2x\right)^{2}}{m^{2}}\right)\right)\right]\nonumber \\
 & +\mathcal{O}\left(\epsilon^{1}\right).
 \label{eq:I2x}
\end{align}
The singular term $\left(1-2x\right)^{-2-2\epsilon}$ can be expressed through the distribution identity:
\begin{align}
\lim_{\epsilon\rightarrow0}\left(\frac{1}{2}-x\right)^{-2-2\epsilon} & \rightarrow\left(-\frac{1}{2\epsilon}-\log2\right)\delta'\left(x-\frac{1}{2}\right)-
2\delta\left(x-\frac{1}{2}\right)+\left[\frac{1}{\left(\frac{1}{2}-x\right)^{2}}\right]_{++},
\label{eq:dis_identy1}
\end{align}
with the single IR pole now manifest.
All distribution identities required in this work have been assembled in Appendix~\ref{Distribution:Identities}.

Substituting (\ref{eq:dis_identy1}) into (\ref{eq:I2x}), and truncating to order $\epsilon^{0}$, we obtain
\begin{align}
 & \mathcal{I}_{2}\left(x\right)\nonumber \\
= & \frac{C_{F}g_{s}^{2}}{4\pi^{2}}\left\{ \frac{1}{8}\delta'\left(x-\frac{1}{2}\right)\left[-x\left(1-2x\right)^{2}-\frac{x\left(1+4x-4x^{2}\right)}{\epsilon}-x\left(1+4x-4x^{2}\right)\ln\left(\frac{\Lambda^{2}}{m^{2}}+\left(1-2x\right)^{2}\right)\right]\right.\nonumber \\
 & \left.+\frac{x\left(1+4x-4x^{2}\right)}{\left[\left(1-2x\right)^{2}\right]_{++}}+\frac{m^{2}x\left(1+4x-4x^{2}\right)}{\Lambda^{2}+m^{2}\left(1-2x\right)^{2}}-\frac{1}{2}\delta\left(x-\frac{1}{2}\right)\right\} \theta\left(x\right)\theta\left(\frac{1}{2}-x\right)+\left(x\rightarrow1-x\right).
\end{align}

Now we have the ultimate result of $\mathcal{I}(x)$:
\begin{align}
\mathcal{I}\left(x\right)= & \mathcal{I}_{1}\left(x\right)+\mathcal{I}_{2}\left(x\right)\nonumber \\
= & \frac{C_{F}g_{s}^{2}}{4\pi^{2}}\left[x\left(\frac{\left(1+4x-4x^{2}\right)}{\left(1-2x\right)^{2}}-\frac{m^{2}\left(1+4x-4x^{2}\right)}{\Lambda^{2}+\left(1-2x\right)^{2}m^{2}}-\frac{\Lambda^{2}}{\Lambda^{2}+\left(1-2x\right)^{2}m^{2}}\right.\right.\nonumber \\
 & \left.\left.+\ln\left(1+\frac{\Lambda^{2}}{m^{2}\left(1-2x\right)^{2}}\right)\right)\theta\left(x\right)\theta\left(\frac{1}{2}-x\right)+\left(x\rightarrow1-x\right)\right]_{++}\nonumber \\
 & +\frac{C_{F}g_{s}^{2}}{4\pi^{2}}\delta\left(x-\frac{1}{2}\right)\left[\frac{1}{4}\left(\frac{1}{\epsilon}+
 \ln\mu_{\rm IR}^{2}\right)+\mathcal{\mathcal{O}}\left(\epsilon^{0}\right)\right],
\label{Ix:final:expression}
\end{align}
which contains the IR pole $\delta\left(x-\frac{1}{2}\right)/\left(4\epsilon\right)$.

We also need include the effects due to the quark wave function
renormalization, as outlined in (\ref{eq:WvFncRnml_1loop}).
The Feynman part of such contributions are
\begin{align}
\delta Z_{q}= & iC_{F}g_{s}^{2}\left(\frac{\mu_{\rm IR}^{2}e^{\gamma_E}}{4\pi}\right)^{\epsilon}\int\frac{d^{4-2\epsilon}k}{\left(2\pi\right)^{4-2\epsilon}}\bar{u}\left(p\right)\gamma^{\mu}\frac{1}{k\!\!\!/+p\!\!\!/-m+i\epsilon}\gamma^{+}\nonumber \\
 & \times\frac{1}{k\!\!\!/+p\!\!\!/-m+i\epsilon}\gamma_{\mu}u\left(p\right)\frac{1}{k^{2}+i\epsilon}
 /\left[\bar{u}\left(p\right)\gamma^{+}u\left(p\right)\right],
\end{align}

\begin{align}
\delta Z_{\bar{q}}= & iC_{F}g_{s}^{2}\left(\frac{\mu_{\rm IR}^{2}e^{\gamma_E}}{4\pi}\right)^{\epsilon}\int\frac{d^{4-2\epsilon}k}{\left(2\pi\right)^{4-2\epsilon}}\bar{v}\left(p\right)\gamma^{\mu}\frac{1}{k\!\!\!/-p\!\!\!/-m+i\epsilon}\gamma^{+}\nonumber \\
 & \times\frac{1}{k\!\!\!/-p\!\!\!/-m+i\epsilon}\gamma_{\mu}v\left(p\right)
 \frac{1}{k^{2}+i\epsilon}/\left[\bar{v}\left(p\right)\gamma^{+}v\left(p\right)\right].
\end{align}
These constants are also IR-divergent:
\begin{align}
\left(\delta Z_{q}+\delta Z_{\bar{q}}\right)\delta\left(x-\frac{1}{2}\right) = & -\frac{C_{F}g_{s}^{2}}{4\pi^{2}}\left[\frac{1}{4}\left(\frac{1}{\epsilon}+\ln\mu_{\rm IR}^{2}\right)+\mathcal{O}\left(\epsilon^{0}\right)\right]\delta\left(x-\frac{1}{2}\right).
\label{wvfrenorm:feyn:part}
\end{align}
Note the $\delta'\left(x-\frac{1}{2}\right)$ have cancelled between
two symmetric pieces under $x\rightarrow1-x$.

It is reassuring that the IR poles exactly
cancel upon summing $\mathcal{I}(x)$ in (\ref{Ix:final:expression} ) and
(\ref{wvfrenorm:feyn:part}).

\subsection{Quasi-DA}

For the quasi-DA of $J/\psi^{\parallel}$, we choose
$n^{\mu}$ as the space-like reference vector as specified
in (\ref{reference:vector:n:def}).
The Feynman part of the vertex diagram in (\ref{eq:DA:one:loop:vertex}) is
\begin{align}
\tilde{\mathcal{I}}\left(x\right)= & \frac{C_{F}}{\left(2\pi\right)^{4-2\epsilon}}\left(\frac{\mu_{IR}^{2}e^{\gamma_E}}{4\pi}\right)^{\epsilon}\int d^{d}k\,\bar{u}\left(p\right)\left(-ig_{s}\gamma^{\mu}\right)
\frac{i}{p\!\!\!/+k\!\!\!/-m+i\epsilon}\gamma^{z}\frac{i}{p\!\!\!/-k\!\!\!/-m+i\epsilon}\nonumber \\
 & \times\left(-ig_{s}\gamma^{\nu}\right)v\left(p\right)\frac{-ig_{\mu\nu}}{k^{2}+i\epsilon}\delta\left(x-\frac{1}{2}
 -\frac{k^{z}}{2p^{z}}\right)/\left[\bar{u}\left(p\right)\gamma^{z}v\left(p\right)\right]\nonumber \\
= & \frac{-ig_{s}^{2}C_{F}}{\left(2\pi\right)^{4-2\epsilon}}\left(\frac{\mu_{\rm IR}^{2} e^{\gamma_E}}{4\pi}\right)^{\epsilon}
\frac{2\pi^{1-\epsilon}}{\Gamma\left(1-\epsilon\right)}\int_{0}^{\Lambda}dk_{\perp}\,k_{\perp}^{1-2\epsilon}\int dk^{0}dk^{z}\nonumber \\
 & \times\frac{\left(2-2\epsilon\right)\left(k^{2}-2k^{z}\left(\frac{p^{z}k^{0}}{p^{0}}-k^{z}\right)\right)
 -4m^{2}}{\left[\left(p+k\right)^{2}-m^{2}+i\epsilon\right]\left[\left(p-k\right)^{2}-m^{2}+i\epsilon\right]} \frac{\delta\left(x-\frac{1}{2}-\frac{k^{z}}{2p^{z}}\right)}{k^{2}+i\epsilon}.
\end{align}
We first perform the $k^{0}$ integration by contour method, then
use the $\delta$-function to trade the $k^{z}$ for the dimensionless
momentum fraction $x$. Nevertheless, since the propagators are
now quadratic in $k^{0}$, the poles are always dispersed in both upper
and lower complex plane, irrespective of the range where $x$ lies in.
After integrating over $k^{0}$ and $k^{z}$, we have
\begin{align}
 & \tilde{\mathcal{I}}\left(x\right)\nonumber \\
= & \frac{g_{s}^{2}C_{F}}{\left(2\pi\right)^{3-2\epsilon}}\frac{4\pi^{1-\epsilon}\mu_{\rm IR}^{2\epsilon}}{\Gamma\left(1-\epsilon\right)}\int_{0}^{\Lambda}dk_{\perp}\,k_{\perp}^{1-2\epsilon}
\frac{p^{z}}{4p^{0}}\nonumber \\
 & \times\left\{ \left[\frac{\left(m^{2}+2\left(p^{z}\right)^{2}x\right)\sqrt{k_{\perp}^{2}+m^{2}+4\left(p^{z}\right)^{2}x}
 \left(\epsilon-1\right)+p^{0}\left(4\left(p^{z}\right)^{2}x^{2}\left(\epsilon-1\right)+m^{2}\epsilon\right)}
 {\sqrt{k_{\perp}^{2}+m^{2}+4\left(p^{z}\right)^{2}x^{2}}\left(m^{2}+2\left(p^{z}\right)^{2}x-p^{0}
 \sqrt{k_{\perp}^{2}m^{2}+4\left(p^{z}\right)^{2}x^{2}}\right)^{2}}\right]_{\text{1st}}\right.\nonumber \\
 & +\left[\frac{-\left(m^{2}\!+\!2\left(p^{z}\right)^{2}\left(\!1\!-\!x\!\right)\right)\sqrt{k_{\perp}^{2}\!+
 \!m^{2}\!+\!4\left(p^{z}\right)^{2}\left(\!1\!-\!x\!\right)^{2}}\left(\epsilon-1\right)+
 p^{0}\left(4\left(p^{z}\right)^{2}\left(\!1\!-\!x\!\right)^{2}\left(\!\epsilon\!-\!1\!\right)\!+
 \!m^{2}\epsilon\right)}{\sqrt{k_{\perp}^{2}+m^{2}+4\left(p^{z}\right)^{2}\left(1-x\right)^{2}}
 \left(m^{2}+2\left(p^{z}\right)^{2}\left(1-x\right)-p^{0}\sqrt{k_{\perp}^{2}+m^{2}+4\left(p^{z}\right)^{2}
 \left(1-x\right)^{2}}\right)^{2}}\right.\nonumber \\
 & \left.\left.-\frac{2\left(m^{2}p^{0}-\left(p^{z}\right)^{2}\left(1-2x\right)
 \left(\sqrt{k_{\perp}^{2}+\left(p^{z}\right)^{2}\left(1-2x\right)^{2}}-
 p^{0}\left(1-2x\right)\right)\left(\epsilon-1\right)\right)}{\sqrt{k_{\perp}^{2}+
 \left(p^{z}\right)^{2}\left(1-2x\right)^{2}}\left(p^{0}\sqrt{k_{\perp}^{2}+
 \left(p^{z}\right)^{2}\left(1-2x\right)^{2}}-p^{z}
 \left(1-2x\right)\right)^{2}}\right]_{\text{2nd}}\right\},
 \label{eq:quasiFey_k0kzInted}
\end{align}
where $p^{0}=\sqrt{\left(p^{z}\right)^{2}+m^{2}}$.

The first piece $\mathcal{\tilde{I}}_{1}(x)$ (labelled by
the subscript ``1st'') in (\ref{eq:quasiFey_k0kzInted}) is IR finite
because its denominator does not vanish as $k_{\perp}\rightarrow 0$
at $x\rightarrow {1\over 2}$. After sending $\epsilon \to 0$, and
performing the $k_{\perp}$ integration, we obtain
\begin{align}
\tilde{\mathcal{I}}_{1}\left(x\right) = & \frac{\left(m^{2}+4\left(p^{z}\right)^{2}x\left(1-x\right)\right)}{8\pi^{2}\left(p^{0}\right)^{2}
\left(1-2x\right)^{2}p^{z}\left(m^{2}\left(p^{z}\right)^{2}\left(1-2x\right)^{2}+\Lambda^{2}
\left(p^{0}\right)^{2}\right)}\nonumber \\
 & \times\left[\left(m^{2}\Lambda^{2}\sqrt{m^{2}+4\left(p^{z}\right)^{2}x^{2}}-p^{0}\right)+\Lambda^{2}
 \left(p^{z}\right)^{2}\left(\sqrt{m^{2}+4\left(p^{z}\right)^{2}x^{2}}-2p^{0}x\right)\right.\nonumber \\
 & \left.+m^{2}\left(p^{z}\right)^{2}\left(1-2x\right)^{2}\left(\sqrt{m^{2}+4\left(p^{z}\right)^{2}x^{2}}-
 \sqrt{\Lambda^{2}+m^{2}+4\left(p^{z}\right)^{2}x^{2}}\right)\right]\nonumber \\
 & +\frac{p^{z}\left(m^{2}+2\left(p^{z}\right)^{2}x\right)}{16\pi^{2}\left(p^{0}\right)^{3}}
 \left[-\ln\left(\frac{m^{2}+2\left(p^{z}\right)^{2}x+p^{0}\sqrt{\Lambda^{2}+m^{2}+
 4\left(p^{z}\right)^{2}x^{2}}}{m^{2}+2\left(p^{z}\right)^{2}x-p^{0}\sqrt{\Lambda^{2}+
 m^{2}+4\left(p^{z}\right)^{2}x^{2}}}\right)\right.\nonumber \\
 & \left.+\ln\left(\frac{m^{2}+2\left(p^{z}\right)^{2}x+p^{0}\sqrt{m^{2}+4\left(p^{z}\right)^{2}x^{2}}}
 {m^{2}+2\left(p^{z}\right)^{2}x-p^{0}\sqrt{m^{2}+4\left(p^{z}\right)^{2}x^{2}}}\right)-
 \ln\left(1+\frac{\Lambda^{2}\left(p^{0}\right)^{2}}{m^{2}\left(p^{z}\right)^{2}
 \left(1-2x\right)^{2}}\right)\right].
\end{align}

The second piece, $\tilde{\mathcal{I}}_{2}\left(x\right)$ (denoted with subscript
``2nd'') in (\ref{eq:quasiFey_k0kzInted}) potentially contains IR
singularity at $x={1\over 2}$. Integrating over the transverse momentum by
brute force, we obtain
\begingroup
\allowdisplaybreaks
\begin{align}
 & \tilde{\mathcal{I}}_{2}
=  \frac{C_{F}g_{s}^{2}e^{\epsilon\gamma}\mu_{\rm IR}^{2\epsilon}}{4\pi^{2}\Lambda^{2\epsilon}}
\left[\frac{\Lambda^{2}\left(m^{2}-2
\left(p^{z}\right)^{2}\left(x-1\right)\right)^{2}\sqrt{m^{2}+4\left(p^{z}\right)^{2}\left(x-1\right)^{2}}}
{2\left(p^{z}\right)^{3}m^{4}\left(1-2x\right)^{4}\Gamma\left(1-\epsilon\right)}\right.\nonumber \\
 & \times F_{1}\left(1-\epsilon,2,-\frac{1}{2},2-\epsilon,-\frac{\left(p^{0}\right)^{2}\Lambda^{2}}{m^{2}
 \left(p^{z}\right)^{2}\left(1-2x\right)^{2}},-\frac{\Lambda^{2}}{m^{2}+4\left(p^{z}\right)^{2}\left(1-x\right)^{2}}\right)\nonumber \\
 & -\frac{\Lambda^{2}\left(p^{z}\right)^{2}\left|1-2x\right|}{m^{4}\left(1-2x\right)^{4}\Gamma
 \left(1-\epsilon\right)}F_{1}\left(1-\epsilon,2,-\frac{1}{2},2-\epsilon,-\frac{\left(p^{0}\right)^{2}
 \Lambda^{2}}{m^{2}\left(p^{z}\right)^{2}\left(1-2x\right)^{2}},-\frac{\Lambda^{2}}{\left(p^{z}\right)^{2}
 \left(1-x\right)^{2}}\right)\nonumber \\
 & +\frac{p^{z}\Lambda^{2\epsilon}\left(p^{0}\right)^{2\epsilon-1}}{\left(1+2\epsilon\right)\Gamma
 \left(1-\epsilon\right)}\frac{m^{2}+\left(p^{z}\right)^{2}\left(1-2x\right)^{2}\left(\epsilon-1\right)}
 {\left(p^{0}\sqrt{\Lambda^{2}+\left(p^{z}\right)^{2}\left(1-2x\right)^{2}}+\left(p^{z}\right)^{2}
 \left(2x-1\right)\right)^{1+2\epsilon}}\nonumber \\
 & \!\times\!F_{1}\!\!\left(\!\!2\epsilon\!+\!1;\epsilon,\epsilon;2\epsilon\!+\!2;\frac{p^{z}
 \left(p^{z}\left(2x\!-\!1\right)\!+\!p^{0}\left|1\!-\!2x\right|\right)}{\left(2x\!-\!1\right)
 \left(p^{z}\right)^{2}\!+\!p^{0}\sqrt{\Lambda^{2}\!+\!\left(p^{z}\right)^{2}\left(1\!-\!2x\right)^{2}}},
 \frac{p^{z}\left(p^{z}\left(2x\!-\!1\right)\!-\!p^{0}\left|1\!-\!2x\right|\right)}{\left(2x\!-\!1\right)
 \left(p^{z}\right)^{2}\!+\!p^{0}\sqrt{\Lambda^{2}\!+\!\left(p^{z}\right)^{2}\left(1\!-\!2x\right)^{2}}}\!\!\right)\nonumber \\
 & -\!\frac{p^{z}\Lambda^{2\epsilon}\left(p^{0}\right)^{2\epsilon-1}}{2\left(1+2\epsilon\right)\Gamma
 \left(1-\epsilon\right)}\frac{m^{2}\epsilon\!+\!4\left(p^{z}\right)^{2}\left(1\!-\!x\right)^{2}
 \left(\epsilon\!-\!1\right)}{\left(p^{0}\!\sqrt{\!\Lambda^{2}\!+\!m^{2}\!+\!4\left(\!p^{z}\!\right)^{2}
 \left(\!1\!-\!x\!\right)^{2}}\!-\!m^{2}\!+\!2\left(\!p^{z}\!\right)^{2}\left(x\!-\!1\right)\right)^{1+2\epsilon}}\nonumber \\
 & \times F_{1}\left(2\epsilon+1,\epsilon,\epsilon,2\epsilon+2,\frac{m^{2}-2\left(p^{z}\right)^{2}
 \left(x-1\right)-p^{0}\sqrt{m^{2}+4\left(p^{z}\right)^{2}\left(x-1\right)^{2}}}{m^{2}-
 2\left(p^{z}\right)^{2}\left(x-1\right)-p^{0}\sqrt{\Lambda^{2}\!\!+\!\!m^{2}+4\left(p^{z}\right)^{2}
 \left(x-1\right)^{2}}},\right.\nonumber \\
 & \left.\frac{m^{2}-2\left(p^{z}\right)^{2}\left(x-1\right)+p^{0}\sqrt{m^{2}+4\left(p^{z}\right)^{2}
 \left(x-1\right)^{2}}}{m^{2}-2\left(p^{z}\right)^{2}\left(x-1\right)-p^{0}\sqrt{\Lambda^{2}+m^{2}+
 4\left(p^{z}\right)^{2}\left(x-1\right)^{2}}}\right)\nonumber \\
 & +\frac{\Lambda^{2}\left(m^{2}+2\left(p^{z}\right)^{2}x\right)}{4m^{2}p^{z}p^{0}
 \left(1-2x\!\right)^{2}\Gamma\left(1-\epsilon\right)}{}_{2}F_{1}\left(1,1-\epsilon;2-
 \epsilon;-\frac{\left(p^{0}\right)^{2}\Lambda^{2}}{m^{2}\left(p^{z}\right)^{2}\left(1-2x\right)^{2}}\right)\nonumber \\
 & +\frac{\Lambda^{2}\left(m^{6}-m^{4}\left(p^{z}\right)^{2}\left(x-1\right)+12m^{2}
 \left(p^{z}\right)^{4}\left(x-1\right)^{2}+2\left(p^{z}\right)^{6}\left(4x^{3}-6x+3\right)
 \right)}{2\left(p^{z}\right)^{3}p^{0}m^{4}\left(1-2x\right)^{4}\Gamma\left(1-\epsilon\right)}\nonumber \\
 & \times{}_{2}F_{1}\!\left(\!1,1-\epsilon,2-\epsilon,-\!\frac{\left(\!p^{0}\!\right)^{2}
 \mu^{2}}{m^{2}\!\left(\!p^{z}\!\right)^{2}\!\left(\!1\!\!-\!\!2x\!\right)^{2}}\!\right)+
 \frac{p^{z}\Gamma\left(2\epsilon\!+\!1\right)}{2\left(\!p^{0}\!\right)^{1-2\epsilon}
 \Gamma\left(\epsilon\!+\!2\right)}\frac{m^{2}\epsilon\!+\!4\left(p^{z}\right)^{2}
 \left(x\!-\!1\right)^{2}\left(\epsilon\!-\!1\right)}{\left(\!\!p^{0}\!\sqrt{\!m^{2}\!\!+
 \!\!4\left(\!p^{z}\!\right)^{2}\!\left(\!1\!\!-\!\!x\!\right)^{2}}\!\!-\!\!m^{2}\!\!+
 \!\!2\left(\!p^{z}\!\right)^{2}\!\left(\!x\!\!-\!\!1\!\right)\!\!\right)^{\!\!1\!+\!2\epsilon}}\nonumber \\
 & \times{}_{2}F_{1}\left(\epsilon,2\epsilon+1,\epsilon+2,\frac{m^{2}-2\left(p^{z}\right)^{2}
 \left(x-1\right)+p^{0}\sqrt{m^{2}+4\left(p^{z}\right)^{2}\left(x-1\right)^{2}}}{m^{2}-
 2\left(p^{z}\right)^{2}\left(x-1\right)-p^{0}\sqrt{m^{2}+4\left(p^{z}\right)^{2}\left(x-1\right)^{2}}}\right)\nonumber \\
 & -\frac{p^{z}\Gamma\left(2\epsilon+1\right)}{\Gamma\left(\epsilon+2\right)\left(p^{0}\right)^{1-2\epsilon}}
 \left(m^{2}+\left(p^{z}\right)^{2}\left(1-2x\right)^{2}\left(\epsilon-1\right)\right)\left(\frac{1}
 {p^{z}\left(p^{0}\left|1-2x\right|+p^{z}\left(2x-1\right)\right)}\right)^{1+2\epsilon}\nonumber \\
 & \left.\times{}_{2}F_{1}\left(\epsilon,2\epsilon+1;\epsilon+2;-\frac{m^{2}+2\left(p^{z}\right)^{2}+2p^{z}
 p^{0}\text{sgn\ensuremath{\left(1-2x\right)}}}{m^{2}}\right)\right],\label{eq:QuasiFeyRes}
\end{align}
\endgroup
where ${}_{2}F_{1}$ is the hypergeometric function, $F_{1}$ the Appell $F_{1}$ function,
and $\text{sgn}(x)$ denotes the sign function.

In $\tilde{\mathcal{I}}_{2}\left(x\right)$, those terms containing
$_{2}F_{1}$ functions can be manipulated according to the method in Sec.~\ref{Appendix:details:lcda},
with the IR pole readily isolated with the aid of distribution identities.
Nevertheless, manipulation of Appell $F_{1}$ functions is much more challenging due to its excessive complication.
Since the IR singularity is always exactly located at
$x={1\over 2}$, we find it beneficial to use the subtraction trick. We
first identify the asymptotic behavior of the Appell $F_{1}$ function near $x\to {1\over 2}$, then apply the distribution identities to isolate the respective IR poles. The difference between the Appell $F_{1}$ and its asymptotic form is IR finite, which is amenable to simple Taylor-expansion in powers of $\epsilon$.

The asymptotic forms of Appell $F_{1}$ functions required in this work
have been tabulated in Appendix.~\ref{AsymptForm:Appell:F1}.

Among all terms containing $F_{1}$ functions in (\ref{eq:QuasiFeyRes}),
only the first one possibly develops an IR singularity.
Following the aforementioned technique, we can identify the IR pole associated with this term:
\begin{align}
\tilde{\mathcal{I}}_{2}^{F_{1}}\left(x\right)= & \frac{C_{F}g_{s}^{2}e^{\epsilon\gamma}\mu_{\rm IR}^{2\epsilon}}{4\pi^{2}\Lambda^{2\epsilon}}\frac{\Lambda^{2}\left(m^{2}-2\left(p^{z}\right)^{2}\left(x-1\right)\right)^{2}\sqrt{m^{2}+4\left(p^{z}\right)^{2}\left(x-1\right)^{2}}}{2\left(p^{z}\right)^{3}m^{4}\Gamma\left(1-\epsilon\right)\left(1-2x\right)^{4}}\nonumber \\
 & \times F_{1}\left(1-\epsilon,-\frac{1}{2},1,2-\epsilon,-\frac{\left(p^{0}\right)^{2}\Lambda^{2}}{m^{2}\left(p^{z}\right)^{2}\left(1-2x\right)^{2}},-\frac{\Lambda^{2}}{m^{2}+4\left(p^{z}\right)^{2}\left(1-x\right)^{2}}\right)\nonumber \\
 & =\frac{C_{F}g_{s}^{2}}{4\pi^{2}}\left[\frac{3p^{z}p^{0}}{16m^{2}}\left(\frac{1}{\epsilon}+\ln\mu_{\rm IR}^{2}\right)+
 \mathcal{O}\left(\epsilon^{0}\right)\right]\delta\left(x-\frac{1}{2}\right).\label{eq:Qua_F1_term}
\end{align}

Summing all the IR-divergent terms in (\ref{eq:QuasiFeyRes}), we obtain
\begin{align}
\left[\tilde{\mathcal{I}_2}(x)\right]_{\frac{1}{\epsilon}}= & \frac{C_{F}g_{s}^{2}}{4\pi^{2}}\delta\left(x-\frac{1}{2}\right)\left[\frac{1}{4}\left(\frac{1}{\epsilon}+\ln\mu_{\rm IR}^{2}\right)+\mathcal{O}\left(\epsilon^{0}\right)\right],
\end{align}
which has the same single-IR pole as in the light-cone case.

The complete result of $\tilde{\mathcal{I}}$ reads
\begin{align}
\tilde{\mathcal{I}}\left(x\right)= & \left[\left.\tilde{\mathcal{I}}_{1}\left(x\right)\right|_{\epsilon=0}+\left.\tilde{\mathcal{I}}_{2}\left(x\right)\right|_{\epsilon=0}\right]_{++}
+\frac{C_{F}g_{s}^{2}}{4\pi^{2}}\left[\frac{1}{4}\left(\frac{1}{\epsilon}+\ln\mu_{\rm IR}^{2}\right)+\mathcal{O}\left(\epsilon^{0}\right)\right]\delta\left(x-\frac{1}{2}\right).\label{eq:Apdx_phi}
\end{align}

We also need include the contribution from the quark self-energy diagrams.
According to (\ref{eq:WvFncRnml_1loop}),
the Feynman part of the quark wave function renormalization
constant is
\begin{align}
\delta\tilde{Z}_{q}= & iC_{F}g_{s}^{2}\left(\frac{\mu_{\rm IR}^{2}e^{\gamma_E}}{4\pi}\right)^{\epsilon}\int\frac{d^{4-2\epsilon}k}{\left(2\pi\right)^{4-2\epsilon}}
\bar{u}\left(p\right)\gamma^{\mu}\frac{1}{k\!\!\!/+p\!\!\!/-m+i\epsilon}\gamma^{z}
\nonumber \\
 & \times\frac{1}{k\!\!\!/+p\!\!\!/-m+i\epsilon}\gamma_{\mu}u\left(p\right)\frac{1}{k^{2}+
 i\epsilon}/\left[\bar{u}\left(p\right)\gamma^{z}u\left(p\right)\right],
\end{align}
\begin{align}
\delta\tilde{Z}_{\bar{q}}= & iC_{F}g_{s}^{2}\left(\frac{\mu_{\rm IR}^{2}e^{\gamma_E}}{4\pi}\right)^{\epsilon}\int\frac{d^{4-2\epsilon}k}{\left(2\pi\right)^{4-2\epsilon}}\bar{v}
\left(p\right)\gamma^{\mu}\frac{1}{k\!\!\!/-p\!\!\!/-m+i\epsilon}\gamma^{z}
\nonumber \\
 & \times\frac{1}{k\!\!\!/-p\!\!\!/-m+i\epsilon}\gamma_{\mu}v\left(p\right)\frac{1}{k^{2}+i\epsilon}/
 \left[\bar{v}\left(p\right)\gamma^{z}v\left(p\right)\right].
\end{align}
Their net contribution is
\begin{align}
\left(\delta\tilde{Z}_{q}+\delta\tilde{Z}_{\bar{q}}\right)\delta\left(x-\frac{1}{2}\right)= & -\frac{C_{F}g_{s}^{2}}{4\pi^{2}}\left[\frac{1}{4}\left(\frac{1}{\epsilon}+\ln\mu_{\rm IR}^{2}\right)+\mathcal{O}\left(\epsilon^{0}\right)\right]\delta\left(x-\frac{1}{2}\right).
\label{eq:Apdx_dZ}
\end{align}

It's straightforward to check the single IR pole
cancels between (\ref{eq:Apdx_phi}) and (\ref{eq:Apdx_dZ}).

\section{Distribution Identities}
\label{Distribution:Identities}

 In DR, the IR divergences usually originate from terms such as $\left|1/2-x\right|^{-1-2\epsilon}$, $\left|1/2-x\right|^{-1-2\epsilon}\ln\left|1/2-x\right|$ and $\left|1/2-x\right|^{-2-2\epsilon}$, {\it etc.}.
 In this work, the following distribution identities have been utilized to express the above singular structures as
the IR pole together with distribution functions:
\begin{subequations}
\begin{eqnarray}
\lim_{\epsilon\rightarrow0}\int_{0}^{\frac{1}{2}}dx\,g\left(x\right)\!\times\!\left|\frac{1}{2}\!-\!x\right|^{-1-2\epsilon}\!\!\! & =\!\!\! & \int_{0}^{\frac{1}{2}}dx\,g\left(x\right)\!\times\!\left\{ \vphantom{\left[\frac{\ln\left(\frac{1}{2}-x\right)}{\frac{1}{2}-x}\right]_{+}}\!\left(\!-\frac{1}{2\epsilon}\!-\!\ln2\!\right)\delta\left(\!x\!-\!\frac{1}{2}\!\right)\!\right\} \nonumber \\
 &  & \left.+\left[\frac{1}{\frac{1}{2}-x}\right]_{+}\right\} ,\\
\lim_{\epsilon\rightarrow0}\int_{\frac{1}{2}}^{1}dx\,g\left(x\right)\times\left|\frac{1}{2}\!-\!x\right|^{-1-2\epsilon} & \!\!\!=\!\!\! & \int_{\frac{1}{2}}^{1}dx\,g\left(x\right)\!\times\!\left\{ \vphantom{\left[\frac{\ln\left(\frac{1}{2}-x\right)}{\frac{1}{2}-x}\right]_{+}}\!\left(\!-\frac{1}{2\epsilon}\!-\!\ln2\!\right)\delta\left(\!x\!-\!\frac{1}{2}\!\right)\right.\nonumber \\
 &  & \left.+\left[\frac{1}{x-\frac{1}{2}}\right]_{+}\right\},
\end{eqnarray}
\end{subequations}
\begin{subequations}
\begin{eqnarray}
\lim_{\epsilon\rightarrow0}\int_{0}^{\frac{1}{2}}dx\,g\left(x\right)\!\times\!\left|\frac{1}{2}\!-\!x\right|^{-1-2\epsilon}\!\!\ln\left|\frac{1}{2}\!-\!x\right| & \!\!\!=\!\!\! & \int_{0}^{\frac{1}{2}}dx\,g\left(x\right)\!\times\!\left\{ \vphantom{\left[\frac{\ln\left(\frac{1}{2}-x\right)}{\frac{1}{2}-x}\right]_{+}}\left[\!-\frac{1}{4\epsilon^{2}}\!+\!\frac{\ln^{2}2}{2}\!\right]\delta\left(\!x-\frac{1}{2}\!\right)\right.\nonumber \\
 &  & \left.+\left[\frac{\ln\left(\frac{1}{2}-x\right)}{\frac{1}{2}-x}\right]_{+}\right\} ,\\
\lim_{\epsilon\rightarrow0}\int_{\frac{1}{2}}^{1}dx\,g\left(x\right)\!\times\!\left|\frac{1}{2}\!-\!x\right|^{-1-2\epsilon}\!\!\ln\left|\frac{1}{2}\!-\!x\right| & \!\!\!=\!\!\! & \int_{\frac{1}{2}}^{1}dx\,g\left(x\right)\!\times\!\left\{ \vphantom{\left[\frac{\ln\left(\frac{1}{2}-x\right)}{\frac{1}{2}-x}\right]_{+}}\left[\!-\frac{1}{4\epsilon^{2}}\!+\!\frac{\ln^{2}2}{2}\!\right]\delta\left(\!x-\frac{1}{2}\!\right)\right.\nonumber \\
 &  & \left.+\left[\frac{\ln\left(x-\frac{1}{2}\right)}{x-\frac{1}{2}}\right]_{+}\right\},
\end{eqnarray}\label{eq:axial_soft}
\end{subequations}
\begin{subequations}
\begin{eqnarray}
\lim_{\epsilon\rightarrow0}\int_{0}^{\frac{1}{2}}dx\,g\left(x\right)\!\times\!\left|\frac{1}{2}\!-\!x\right|^{-2-2\epsilon} & \!\!\!\!=\!\!\! & \int_{0}^{\frac{1}{2}}dx\,g\left(x\right)\!\times\!\left\{ \vphantom{\left[\frac{1}{\left(\frac{1}{2}-x\right)^{2}}\right]_{++}}\!\left(\!-\frac{1}{2\epsilon}\!-\!\ln2\!\right)\delta'\left(\!x\!-\!\frac{1}{2}\!\right)\!\right.\nonumber \\
 &  & \left.-\!2\delta\left(\!x\!-\!\frac{1}{2}\!\right)+\!\!\left[\!\frac{1}{\left(\frac{1}{2}\!-\!x\right)^{2}}\!\right]_{++}\right\} ,\\
\lim_{\epsilon\rightarrow0}\int_{\frac{1}{2}}^{1}dx\,g\left(x\right)\!\times\!\left|\frac{1}{2}\!-\!x\right|^{-2-2\epsilon} & \!\!\!\!=\!\!\! & \int_{\frac{1}{2}}^{1}dx\,g\left(x\right)\!\times\!\left\{ \vphantom{\left[\frac{1}{\left(\frac{1}{2}-x\right)^{2}}\right]_{++}}\!\left(\!\frac{1}{2\epsilon}\!+\!\ln2\!\right)\delta'\left(\!x\!-\!\frac{1}{2}\!\right)\!\right.\nonumber \\
 &  & \left.-\!2\delta\left(\!x\!-\!\frac{1}{2}\!\right)+\!\!\left[\!\frac{1}{\left(\frac{1}{2}\!-\!x\right)^{2}}\!\right]_{++}\right\} ,\label{eq:DisIdnty}
\end{eqnarray}
\end{subequations}
The IR divergence is represented by $\epsilon^{-n}$. The above distribution identities should be understood to be
convolved with a test function $g(x)$ that is regular at $x={1\over 2}$.
The double pole $\epsilon^{-2}$ in (\ref{eq:axial_soft}) stems from the coupled soft and light-cone (axial) singularity.

\section{Asymptotic form of Appell $F_{1}$ functions and pole structures}
\label{AsymptForm:Appell:F1}

When computing the one-loop corrections to the quasi-DAs, we have encountered numerous Appell $F_1$ functions.
Here we present the asymptotic form of
the encountered Appell $F_{1}$ functions near $x={1\over 2}$:
\begin{subequations}
\begin{align}
 & F_{1}\left(1-\epsilon,-\frac{1}{2},1,2-\epsilon,-\frac{\left(p^{0}\right)^{2}\Lambda^{2}}{m^{2}\left(p^{z}\right)^{2}\left(1-2x\right)^{2}},-\frac{\Lambda^{2}}{m^{2}+4\left(p^{z}\right)^{2}\left(1-x\right)^{2}}\right)\nonumber \\
\rightarrow & \left(1-2x\right)^{4}\left[\frac{\Gamma\left(2-\epsilon\right)\Gamma\left(1+\epsilon\right)}{\Gamma\left(1-\epsilon\right)}\left(\frac{p^{0}\Lambda}{p^{z}m}\right)^{2\epsilon-2}\left|1-2x\right|^{-2-2\epsilon}-\frac{1-\epsilon}{\left(1+\epsilon\right)\Gamma\left(1-\epsilon\right)}\left(\frac{p^{z}m}{p^{0}\Lambda}\right)^{4}\right],\\
 & F_{1}\left(1-\epsilon,-\frac{1}{2},1,2-\epsilon,-\frac{\Lambda^{2}}{\left(p^{z}\right)^{2}\left(1-2x\right)^{2}},-\frac{\left(p^{0}\right)^{2}\Lambda^{2}}{m^{2}\left(p^{z}\right)^{2}\left(1-2x\right)^{2}}\right)\nonumber \\
\rightarrow & \left(1\!-\!2x\right)^{3}\!\left[\frac{m^{2}\Lambda}{\left(p^{0}\right)^{2}p^{z}}\frac{2\left(p^{z}\right)^{2}\left(1\!-\!\epsilon\right)}{\left(1\!-\!2\epsilon\right)\mu^{2}}\left(\!1\!-\!2x\!\right)^{-2}\!+\!\frac{\Gamma\!\left(\!2\!-\!\epsilon\!\right)\!\Gamma\!\left(\!\epsilon\!-\!\frac{1}{2}\!\right)}{\sqrt{\pi}}\left(\!\frac{p^{z}}{\Lambda}\!\right)^{3-2\epsilon}\right.\nonumber \\
 & \times\left._{2}F_{1}\left(\!1,-\frac{1}{2}\!+\!\epsilon,\frac{1}{2},\left(\!\frac{p^{z}}{p^{0}}\!\right)^{2}\!\right)\!\left|1-2x\right|^{-1-2\epsilon}\right],\\
 & F_{1}\left(1-\epsilon,-\frac{1}{2},1,2-\epsilon,-\frac{\Lambda^{2}}{m^{2}+4\left(p^{z}\right)^{2}x^{2}},-\frac{\left(p^{0}\right)^{2}\Lambda^{2}}{m^{2}\left(p^{z}\right)^{2}\left(1-2x\right)^{2}}\right)\nonumber \\
\rightarrow & \left(\!1\!-\!2x\!\right)^{3}\left[\frac{m^{2}\left(\!p^{z}\!\right)^{2}}{\left(\!p^{0}\!\right)^{2}
\Lambda^{2}}\left(\!-\frac{1\!-\!\epsilon}{\epsilon}{}_{2}F_{1}
\left(\!-\frac{1}{2},-\epsilon,1\!-\!\epsilon,-\frac{\Lambda^{2}}{\left(\!p^{0}\!\right)^{2}}
\!\right)+\frac{\left(p^{z}\Lambda\right)^{2}}{\left(p^{0}\right)^{4}}{}_{2}F_{1}\left(\!
\frac{1}{2},1\!-\!\epsilon,2\!-\!\epsilon-\frac{\Lambda^{2}}{\left(\!p^{0}\!\right)^{2}}\!\right)\!\right)
\left|1\!-\!2x\right|^{-1}\right.\nonumber \\
 & \left.+\left(\frac{mp^{z}}{\Lambda p^{0}}\right)^{2-2\epsilon}\left|1-2x\right|^{-1-2\epsilon}\Gamma\left(2-\epsilon\right)\Gamma\left(\epsilon\right)\right].
\end{align}
\end{subequations}
Although these Appell $F_{1}$ functions vanish at $x={1\over 2}$, they are usually accompanied with
 $\left(1-2x\right)^{-n}$ ($n=3,4$).
Subtraction algorithm can be applied to isolate their divergent pieces from finite ones.

We take (\ref{eq:Qua_F1_term}) as an example,
\begin{align}
\tilde{\mathcal{I}}_{2}^{F_{1}}\left(x\right)= & \frac{C_{F}g_{s}^{2}e^{\epsilon\gamma_E}\mu_{\rm IR}^{2\epsilon}}{4\pi^{2}\Lambda^{2\epsilon}}\frac{\Lambda^{2}\left(m^{2}-
2\left(p^{z}\right)^{2}\left(x-1\right)\right)^{2}\sqrt{m^{2}+4\left(p^{z}\right)^{2}
\left(x-1\right)^{2}}}{2\left(p^{z}\right)^{3}m^{4}\Gamma\left(1-\epsilon\right)\left(1-2x\right)^{4}}\nonumber \\
 & \times F_{1}\left(1-\epsilon,-\frac{1}{2},1,2-\epsilon,-\frac{\left(p^{0}\right)^{2}\Lambda^{2}}{m^{2}
 \left(p^{z}\right)^{2}\left(1-2x\right)^{2}},
 -\frac{\Lambda^{2}}{m^{2}+4\left(p^{z}\right)^{2}\left(1-x\right)^{2}}\right)\nonumber \\
= & \tilde{\mathcal{I}}_{2}^{F_{1}}(x)\bigg|_{\rm asym}+ \left.\left\{ \tilde{\mathcal{I}}_{2}^{F_{1}}\left(x\right)-\left[\tilde{\mathcal{I}}_{2}^{F_{1}}\left(x\right)\right]_{\text{asym}}\right\} \right|_{\epsilon=0}+\mathcal{O}\left(\epsilon^{1}\right)
\end{align}
where the first term
denotes the asymptotic form of $\tilde{\mathcal{I}}_{2}^{F_{1}}\left(x\right)$, the second term denotes the subtracted part.
The subtracted part is regular at $x={1\over 2}$, thereby one can simply set $\epsilon\to 0 $ in it.

The asymptotic part reads:
\begin{align}
\tilde{\mathcal{I}}_{2}^{F_{1}}(x)\bigg|_{\rm asym} & = \frac{C_{F}g_{s}^{2}e^{\epsilon\gamma}\mu_{\rm IR}^{2\epsilon}}{4\pi^{2}\Lambda^{2\epsilon}}
\frac{\Lambda^{2}\left(m^{2}-2\left(p^{z}\right)^{2}\left(x-1\right)\right)^{2}
\sqrt{m^{2}+4\left(p^{z}\right)^{2}\left(x-1\right)^{2}}}{2\left(p^{z}\right)^{3}m^{4}\Gamma\left(1-\epsilon\right)}\nonumber \\
 & \times\left(\frac{\Gamma\left(2-\epsilon\right)\Gamma\left(1+\epsilon\right)}{\Gamma\left(1-\epsilon\right)}\left(\frac{p^{0}\Lambda}{p^{z}m}\right)^{2\epsilon-2}\left(1-2x\right)^{-2-2\epsilon}-\frac{1-\epsilon}{\left(1+\epsilon\right)\Gamma\left(1-\epsilon\right)}\left(\frac{p^{z}m}{p^{0}\Lambda}\right)^{4}\right)
\end{align}
Rewriting $\left(1-2x\right)^{2+2\epsilon}$ through the distribution
identity in (\ref{eq:DisIdnty}) leads to
\begin{align}
\tilde{\mathcal{I}}_{2}^{F_{1}}\left(x\right)\bigg|_{\rm asym} = & \frac{C_{F}g_{s}^{2}e^{\epsilon\gamma}\mu_{\rm IR}^{2\epsilon}}{4\pi^{2}\Lambda^{2\epsilon}}\frac{\Lambda^{2}\left(m^{2}-2
\left(p^{z}\right)^{2}\left(x-1\right)\right)^{2}\sqrt{m^{2}+4\left(p^{z}\right)^{2}\left(x-1\right)^{2}}}
{2\left(p^{z}\right)^{3}m^{4}\Gamma\left(1-\epsilon\right)}\nonumber \\
 & \times\left(\!\frac{\Gamma\left(2-\epsilon\right)\Gamma\left(1+\epsilon\right)}
 {\Gamma\left(1-\epsilon\right)}\left(\frac{p^{0}\Lambda}{mp^{z}}\right)^{2\epsilon-2}
 \frac{1}{2^{2+2\epsilon}}\left(\vphantom{\left[\!\frac{1}{\left(\frac{1}{2}\!-\!x\right)^{2}}
 \!\right]_{++}}\!\left(\!-\frac{1}{2\epsilon}-\log2\!\right)\delta'\left(x\!-\!\frac{1}{2}\right)\!\right.\right.\nonumber \\
 & \left.\left.-\!2\delta\left(x\!-\!\frac{1}{2}\right)\!+\!\left[\!\frac{1}
 {\left(\frac{1}{2}\!-\!x\right)^{2}}\!\right]_{++}\right)-\!\frac{1-\epsilon}
 {\left(1\!+\!\epsilon\right)\Gamma\left(1\!-\!\epsilon\right)}\frac{m^{4}
 \left(p^{z}\right)^{4}}{\left(p^{0}\right)^{4}\Lambda^{4}}\right)
\end{align}
Therefore, the IR pole of $\tilde{\mathcal{I}}_{2}^{F_{1}}$ becomes
\begin{align}
\left(\tilde{\mathcal{I}}_{2}^{F_{1}}\right)_{\frac{1}{\epsilon}}= & \frac{C_{F}g_{s}^{2}}{4\pi^{2}}\left(\frac{3p^{z}p^{0}}{16m^{2}}\right)\left(\frac{1}{\epsilon}+\ln\mu_{\rm IR}^{2}\right)\delta\left(x-\frac{1}{2}\right).
\end{align}

\end{document}